\documentclass[12pt]{article}
\usepackage[T1]{fontenc}
\usepackage[utf8]{inputenc}
\usepackage{lmodern,amsmath,amssymb,amsthm,mathtools,bm}
\usepackage[a4paper,margin=28mm]{geometry}
\usepackage{microtype,booktabs,graphicx,enumitem,setspace}
\usepackage[authoryear,round]{natbib}
\usepackage[hidelinks,pdftitle={Weak Identification in Bunching Designs: Rare Events and Counterfactual Uncertainty},pdfauthor={Kun Huang}]{hyperref}
\usepackage[nameinlink,noabbrev]{cleveref}
\numberwithin{equation}{section}
\newtheorem{theorem}{Theorem}
\newtheorem{proposition}[theorem]{Proposition}
\newtheorem{lemma}[theorem]{Lemma}
\newtheorem{corollary}[theorem]{Corollary}
\theoremstyle{definition}
\newtheorem{assumption}{Assumption}

\theoremstyle{remark}

\newcommand{\R}{\mathbb R}
\newcommand{\E}{\mathbb E}
\newcommand{\Prob}{\mathbb P}
\newcommand{\1}{\mathbf 1}
\newcommand{\dd}{\,\mathrm d}
\newcommand{\ThetaS}{[a,T]}
\newcommand{\Holder}{\mathcal F_s(c,C,L,\varepsilon)}
\newcommand{\diam}{\operatorname{diam}}
\newcommand{\TV}{\operatorname{TV}}
\newcommand{\Pois}{\operatorname{Pois}}

\newcommand{\clip}{\operatorname{proj}}
\newcommand{\MCReplications}{675000}
\newcommand{\MCDesigns}{135}
\newcommand{\MCRepsPerDesign}{5000}
\newcommand{\HonestCoverageMin}{0.9950}
\newcommand{\WeakWaldMin}{0.8512}
\newcommand{\WeakWaldMax}{0.8718}
\newcommand{\VeryWeakWaldMin}{0.1732}
\newcommand{\VeryWeakWaldMax}{0.1914}
\newcommand{\MaxCoverageMCSE}{0.0071}
\newcommand{\NominalCoverageMCSE}{0.0031}

\newcommand{\PairDesigns}{12}
\newcommand{\PairReplications}{60000}
\newcommand{\PairJointCoverageMin}{0.9952}
\newcommand{\AllReplications}{960000}

\title{Weak Identification in Bunching Designs:\\ Rare Events and Counterfactual Uncertainty}
\author{Kun Huang\thanks{Economics and Management School, Wuhan University, Wuhan, China. Email: \href{mailto:huangkun123huang@163.com}{huangkun123huang@163.com}.}}
\date{October 4, 2026}
\begin{document}
\maketitle
\begin{abstract}
We study shrinking tax kinks when the counterfactual density is locally H\"older continuous and may be accompanied by a known background atom at the threshold. Small policy incentives create two distinct obstacles: few behavioral bunchers and an unobserved relocation interval. For the complete observed distribution, we establish matching minimax bounds for elasticity estimation and honest confidence-set diameter. With kink size $d$, background mass $\gamma$, and smoothness $0<s\leq1$, the rate is $\min\{1,\sqrt{(\gamma+d)/(nd^2)}+d^s\}$. The stochastic component has a rare-event scale when the background mass is of the kink order or smaller and a conventional weak-signal scale when it remains positive. The extrapolation component follows from observationally equivalent counterfactual densities, rather than sampling error. We derive normal and Poisson limits for a conventional bunching ratio, show a coverage failure of its Wald interval, and give finite-sample confidence sets by exact binomial inversion with a deterministic smoothness allowance. The procedure estimates the counterfactual density from an unaffected bin and covers the entire population identified set uniformly, including zero policy incentives. A likelihood-ratio calculation establishes a finite-parameter Poisson limit for a smooth, known counterfactual submodel. Structural Monte Carlo experiments document both coverage protection and its cost in set length. The results apply to the stated exact-atom observation model; general rounding and optimization frictions require additional assumptions.
\end{abstract}
\newpage
\section{Introduction}
An elasticity inferred from bunching is obtained by comparing a behavioral mass with the distribution of people who could have moved to the threshold. The comparison depends on the incentive change, the density of counterfactual choices, and the observation mechanism. A small derivative of bunching mass with respect to elasticity does not, by itself, describe a weakly informative experiment. The variance of the mass can shrink at the same time. Conversely, a precisely measured atom can contain little behavioral information if most of it would have been present without the policy.

This paper studies this distinction in a transparent kink model. Let $d$ be the change in log net-of-tax slopes and $\theta$ an elasticity. Individuals with counterfactual log income between zero and $d\theta$ choose the threshold. If their density is $f$, the behavioral mass is $\int_0^{d\theta}f(x)\dd x$, and its derivative is $df(d\theta)$. When $d$ decreases, both the derivative and the number of bunchers decrease. With no background atom and a positive density at the cutoff, the variance of the empirical mass is of order $d/n$. The relevant stochastic sensitivity is therefore of order $\sqrt{nd}$, rather than $\sqrt n d$. When a fixed background mass is observationally indistinguishable from behavioral bunching, the variance no longer vanishes with $d$, and the latter normalization becomes relevant.

There is also a population identification issue. A H\"older restriction on the counterfactual density does not point identify elasticity at a fixed kink. The density inside the relocation interval is not observed. Changing elasticity and changing that density can leave the atom and both continuous sides of the observed distribution exactly unchanged. We quantify this ambiguity as the kink shrinks. It has order $d^s$ for a class with known smoothness exponent $s\in(0,1]$ and known radius. Shrinking incentives reduce this extrapolation ambiguity while increasing sampling uncertainty. The resulting tradeoff differs from the familiar argument that a larger policy discontinuity necessarily improves structural identification.

Our main result combines these two obstacles for the \emph{complete} mixed observation distribution. Write $\gamma$ for a known background atom, with $\gamma$ bounded below one. The minimax absolute-error risk, and the smallest possible worst-case expected diameter of an honest confidence set, have order
\begin{equation}\label{eq:intro-rate}
 \rho_{n,d,\gamma}=\min\left\{1,\sqrt{\frac{\gamma+d}{nd^2}}+d^s\right\}.
\end{equation}
The constants depend on the density class and parameter space. The bound includes a lower bound for the full sample, not merely a lower bound after discarding continuous observations. A smooth known-density submodel gives the stochastic lower bound. A separate construction produces two admissible densities and two elasticities with exactly the same observed law; it gives the extrapolation lower bound. A regularized ratio and an exact test inversion attain the rate. Thus the density-estimation term can be accommodated without introducing an additional dominant nonparametric rate in this model.

The second contribution concerns distributions and inference across regimes. Without a background atom, a bounded expected number of behavioral bunchers leads to a Poisson ratio limit. With too few bunchers, the conventional ratio concentrates at zero despite a positive elasticity. A plug-in Wald interval collapses to a singleton whenever the bunching count is zero. Its coverage is consequently bounded above by the probability of observing at least one buncher. We also establish convergence of finite collections of full-sample likelihood ratios to Poisson likelihood ratios in a smooth, known-density submodel. This calculation distinguishes a count approximation from an assertion about the full experiment. We do not claim uniform asymptotic equivalence of all nuisance-parameter experiments.

The third contribution is a feasible finite-sample inference procedure. An unaffected bin below the threshold estimates the counterfactual density. Exact binomial intervals for its probability and for the total atom probability are combined with a H\"older bridge over the unobserved interval. Inverting these inequalities avoids division by an estimated structural derivative. The construction covers the population identified set with probability at least the nominal level, uniformly over kink sizes, background masses, and the stated density class. It remains valid for zero incentives, small expected counts, and an empty observed density bin. Its price is conservativeness and, in weak designs, wide or disconnected sets. The procedure does not learn the smoothness radius automatically.

These contributions have a deliberately limited scope. General identification, density restrictions, small-kink approximations, and test inversion in bunching designs are established parts of the literature. In particular, \citet{song2025} already develops bias-aware inference with estimated counterfactual densities. Our claim concerns the matching full-observation bounds and the transition between rare-atom and background-atom experiments under shrinking incentives. The theory is not obtained by merely attaching a generic Gaussian weak-identification confidence set to a bunching moment: the changing variance, the Poisson regime, and the exact observational-equivalence lower bound have to be derived from the relocation model. Nevertheless, the probability and inference tools used in the derivations are classical.

We report structural simulations rather than an empirical application. The simulated observations result from individual isoelastic optimization under a continuous kinked budget. We vary sample size, incentives, counterfactual curvature, nuisance tuning, and background atoms. Every numerical table and figure is generated from executed code. The simulations evaluate both coverage and length; they are not evidence that the chosen smoothness restriction describes administrative data. The practical discussion separates exact behavioral atoms, calibrated background atoms, and more general coarsening mechanisms.

\section{Related literature and scope of the contribution}
\citet{saez2010} relates excess mass at tax kinks to compensated reported-income elasticities. \citet{chetty2011} studies adjustment costs and uses counterfactual density fitting in Danish tax data. The kink and notch distinction, and the role of frictions, are reviewed by \citet{kleven2016}; \citet{kleven2013} uses notches and missing mass to study frictions and structural responses. Our canonical environment is a homogeneous-elasticity kink with exact choices. No notch identification theorem or optimization-friction result is asserted here.

The population counterfactual problem is central to \citet{blomquist2021}. With an unrestricted preference distribution, observing the full taxable-income distribution does not identify the elasticity; restrictions across the kink are needed. \citet{bertanha2023} develops nonparametric bounds and semiparametric methods, including sharp bounds under a Lipschitz density restriction. Their fixed-kink restrictions and bounds are not new results of this paper. \citet{goff2026} studies counterfactual-choice distributions, heterogeneous responses, and shape restrictions. Its latest public revision discusses small-kink limits as well as inference. A small-kink approximation or the collapse of certain population bounds is therefore not, alone, a novel contribution.

The closest inference paper is \citet{song2025}. It identifies parameters under analytic counterfactual densities, relaxes point identification through inequality restrictions, and develops a generalized polynomial strategy with bias-aware test inversion. Its uniform asymptotic result imposes a positive lower bound on a conditional weighted relocation distance (Assumption 9(ii), in the locked full-text version). A pure kink with $d\theta\to0$ does not satisfy that lower bound. Our shrinking experiment uses a H\"older class without analytic continuation, proves a full-observation risk bound, and retains exact binomial validity when the atom becomes rare. This is a difference in the covered experiment and in the quantified precision bound, not an invention of counterfactual estimation or moment inversion.

The weak-identification literature supplies the broader motivation. \citet{staiger1997} studies weak instrumental variables; \citet{stock2000} provides weak generalized method of moments (GMM) asymptotics and robust confidence sets. \citet{andrews2012} separates weak, semi-strong, and strong identification for extremum estimators. \citet{moreira2003} develops conditional likelihood-ratio inference in a structural linear model. \citet{dufour1997} gives general impossibility results when parameters approach nonidentification. These papers do not license a normal approximation for a bounded number of atom observations. Our likelihood and variance calculations identify the appropriate limit in the stated model.

\citet{kaji2021} formulates weak identification in semiparametric models and introduces weak efficiency. \citet{wang2025} treats many weak moments with estimated nuisance parameters. We do not extend either general theory. In our model, the density first step is a bin probability, its error is multiplied by the relocation length, and exact simultaneous probability bounds eliminate the need for a Gaussian nuisance expansion. This simple first step is chosen to attain the minimax order in the specified class. It does not establish that histogram estimation is efficient under richer identifying restrictions.

Honest inference under smoothness restrictions has a separate, established history. \citet{armstrong2018} derives optimal confidence intervals for regression functionals under convex function classes. Exact binomial limits originate with \citet{clopper1934}. We use their coverage principle, a union bound, and an explicit extrapolation inequality. These components are individually classical. The bunching-specific result is that their combination matches lower bounds arising from the complete relocation distribution and from a population-equivalence construction. Local polynomial density estimation, such as \citet{cattaneo2020}, is a useful alternative first step, but substituting it here would require a separate finite-sample or uniformly justified bias analysis.

Finally, \citet{dutz2026} studies kinked budget sets with income effects and heterogeneous elasticities. Its identifying inputs include the kinked earnings distribution and an observed or estimated portion of the no-kink earnings distribution; its application estimates the latter by extrapolation. Our one-schedule experiment instead leaves the local counterfactual unrestricted within a fixed H\"older class. \citet{caetano2025} studies causal responses using bunching in a continuous treatment and outcome variation, an additional source of information absent here. None of our bounds restricts what these richer information sets can identify.

Our literature audit records versions, primary verification sources, and unresolved search limits. In particular, a previously announced conference project on bunching inference could not be obtained as a public full text, and a source-date inconsistency exists between the webpage and downloaded full text of Song's public version. The citation in this paper is locked to the verified full text and arXiv metadata. A failure to find a matching theorem is evidence for the stated comparison, not a proof that no unpublished work has addressed it.

\section{A canonical model and the observed experiment}\label{sec:model}
\subsection{Individual optimization and relocation}
Let gross income be $z>0$, productivity be $A>0$, and the elasticity be $\theta>0$. Utility is
\begin{equation}\label{eq:utility}
 u(c,z;A,\theta)=c-\frac{\theta}{1+\theta}A^{-1/\theta}z^{1+1/\theta}.
\end{equation}
The budget is continuous at a known income threshold $k>0$. Its net-of-tax slopes are $s_0>s_1>0$ below and above $k$. On a linear segment, the first-order condition is $s_j=A^{-1/\theta}z^{1/\theta}$, giving $z_j=A s_j^\theta$. The objective is strictly concave in $z$, and the slope of consumption falls at $k$. It follows that the chosen income is $z_0$ below $k$, $k$ when $z_1\leq k\leq z_0$, and $z_1$ above $k$. Transfers affect the utility level but not these choices because consumption enters linearly.

Define
\begin{equation}\label{eq:transformation}
 X=\log(A s_0^\theta/k),\qquad Y=\log(z/k),\qquad d=\log(s_0/s_1)\geq0.
\end{equation}
The counterfactual log-income variable $X$ has density $f$. Its distribution is a nuisance parameter; the associated productivity distribution is obtained by inverting \eqref{eq:transformation}. Holding a counterfactual density fixed across candidate elasticities does not hold the productivity distribution fixed. That distinction is part of the semiparametric experiment.

The exact choice rule is
\begin{equation}\label{eq:choice}
 Y=T_{\theta,d}(X)=
 \begin{cases}X,&X<0,\\0,&0\leq X\leq d\theta,\\X-d\theta,&X>d\theta.\end{cases}
\end{equation}
The threshold atom has mass $B_{\theta,d}(f)=\int_0^{d\theta}f(x)\dd x$. Left-side choices are unaffected. Right-side choices are translated by the endogenous relocation length $d\theta$. The rule produces an atom and a deformation of support from optimization; it is not an arbitrary density bump.

\subsection{Background atoms}
The separation of background bunchers from policy responses is already explicit in \citet{goff2021}, which conditions on the complementary group and uses auxiliary information about background mass. The mixture below fixes one transparent observation mechanism for a joint precision analysis; the decomposition itself is not a new identification claim.
Let $D$ be independent of $X$ with $\Prob(D=1)=\gamma$. The recorded variable equals zero when $D=1$, and otherwise equals the choice in \eqref{eq:choice}. Write this recorded variable again as $Y$. Here $\gamma$ is a calibrated, policy-independent background probability. This observation model could represent a mixture of threshold reports and accurately recorded optimizing choices. It is \emph{not} a model of arbitrary rounding of every observation or of diffuse adjustment errors. The principal model includes $\gamma=0$.

The full law $P_{\theta,f,d,\gamma}$ has an atom and two densities:
\begin{align}\label{eq:law}
 p&=\Prob(Y=0)=\gamma+(1-\gamma)\int_0^{d\theta}f(x)\dd x,\\
 g_Y(y)&=(1-\gamma)f(y)\quad(y<0),\qquad
 g_Y(y)=(1-\gamma)f(y+d\theta)\quad(y>0).\nonumber
\end{align}
We observe $n$ independent draws from this complete law. The lower-bound experiment includes all their exact continuous values. Our feasible upper bounds use only two summaries used by the selected procedure; they are not assumed sufficient for the complete model.

\begin{assumption}[Local density class]\label{ass:class}
Fix $0<a<T<\infty$, $0<s\leq1$, $L>0$, $0<c<1/4<C<\infty$, $0<\varepsilon\leq1$, and $0\leq\bar\gamma<1$. The parameter is $\theta\in\ThetaS$ and $\gamma\in[0,\bar\gamma]$. The class $\Holder$ contains all Lebesgue densities on $\R$ such that, for $x,y\in[-\varepsilon,\varepsilon]$,
\[
 c\leq f(x)\leq C,\qquad |f(x)-f(y)|\leq L|x-y|^s.
\]
Outside this neighborhood there is no shape restriction. The kink obeys $0\leq d\leq\varepsilon/\max\{1,T\}$, and the chosen bin width obeys $0<h\leq\varepsilon$.
\end{assumption}
The positive density bound excludes a disappearing pool of potential movers, the positive elasticity bound excludes a boundary problem, and $\bar\gamma<1$ ensures that a nonvanishing fraction is accurately recorded. The numerical normalization $c<1/4<C$ lets us use an explicit least-favorable density; it is not an economic restriction on the units of raw income. The smoothness restriction concerns log counterfactual income. Its radius and exponent are inputs to honest inference, not estimates with asserted automatic validity.

For fixed $n,d,\gamma$, the model class is
\[
 \mathcal P_{n,d,\gamma}=\{P_{\theta,f,d,\gamma}^{\otimes n}:\theta\in\ThetaS,\ f\in\Holder\}.
\]
Triangular sequences may change $d$, $\gamma$, $\theta$, and $f$ with $n$, while keeping all constants in \cref{ass:class} fixed. Uniform statements use the infimum or supremum over these laws at each $n$. When $d=0$, the observed law depends on $f$ and $\gamma$ but not on $\theta$.

\subsection{Moments and the density first step}
Let $g=(1-\gamma)f$ and set
\begin{equation}\label{eq:bin}
 \begin{aligned}
 q&=\Prob(-h\leq Y<0)=\int_{-h}^0g(x)\dd x,\qquad \mu_h=q/h,\\
 N_0&=\sum_{i=1}^n\1\{Y_i=0\},\qquad N_-=\sum_{i=1}^n\1\{-h\leq Y_i<0\}.
 \end{aligned}
\end{equation}
The two counts and their complement have a multinomial distribution. In particular, $\operatorname{Cov}(N_0/n,N_-/n)=-pq/n$. Define $\widehat p=N_0/n$, $\widehat q=N_-/n$, and $\widehat\mu_h=\widehat q/h$. Then
\begin{equation}\label{eq:density-rate}
 \E(\widehat\mu_h-\mu_h)^2=\frac{q(1-q)}{nh^2}=O((nh)^{-1}),\qquad
 |\mu_h-g(0)|\leq(1-\gamma)Lh^s/(s+1).
\end{equation}
The exact structural moment is
\begin{equation}\label{eq:exact-moment}
 M(\vartheta,g)=p-\gamma-\int_0^{d\vartheta}g(x)\dd x.
\end{equation}
It vanishes at the true parameter. The local-flat sample moment replaces the integral by $d\vartheta\widehat\mu_h$. Its nuisance error is multiplied by $d\vartheta$. Therefore the density's stochastic rate $\sqrt{nh}$ cannot simply be assigned, without this multiplier, to the structural moment.

\section{Identification, extrapolation, and sensitivity}\label{sec:id}
\subsection{Oracle identification and full-model identification}
For a known $f$, the map $\theta\mapsto B_{\theta,d}(f)$ is strictly increasing when $d>0$. Its oracle derivative is
\begin{equation}\label{eq:oracle-j}
 J_f(\theta,d)=df(d\theta),\qquad
 (1-\gamma)J_f(\theta,d)\in[(1-\bar\gamma)cd,Cd].
\end{equation}
Consequently the atom identifies $\theta$ conditionally on a known counterfactual density. This statement is distinct from point identification in the full model, where $f$ can vary with the parameter. A positive oracle derivative does not remove nuisance directions that offset a change in elasticity.

For an observed population law $P$ at known $(d,\gamma)$, define its identified set by
\begin{equation}\label{eq:identifiedset}
 I(P)=\{\vartheta\in\ThetaS:\text{there is }\widetilde f\in\Holder
 \text{ with }P_{\vartheta,\widetilde f,d,\gamma}=P\}.
\end{equation}
Point identification means $I(P)$ is a singleton. Partial identification means it contains several values. At $d=0$, $I(P)=\ThetaS$ for every compatible law. Weak information along a shrinking sequence is a further concept: even if oracle point identification holds at every positive $d_n$, the sample may fail to concentrate uniformly around the parameter. The following result quantifies a separate full-model ambiguity.

\begin{theorem}[Full-law observational equivalence]\label{thm:equiv}
Under \cref{ass:class}, there are $d_0>0$ and $\kappa>0$ such that, for every $0<d\leq d_0$ and every $\gamma\in[0,\bar\gamma]$, two pairs $(\theta_1,f_1)$ and $(\theta_2,f_2)$ in the class satisfy
\[
 \theta_2-\theta_1=\kappa d^s,\qquad
 P_{\theta_1,f_1,d,\gamma}=P_{\theta_2,f_2,d,\gamma}.
\]
The equality includes the atom and both continuous sides. Therefore no estimator has uniformly vanishing error at a fixed such $d$, and the population identified set can have diameter at least $\kappa d^s$.
\end{theorem}
The proof, in \cref{app:equiv}, changes the density over the unobserved relocation interval and translates its upper continuation. It maintains total probability, positivity, the same H\"older radius, and the same complete observed law. This is stronger than equating two bunching masses. It is also a quantitative local extension of an established identification concern, not a claim that counterfactual nonidentification was previously unknown.

\subsection{A bridge from an unaffected bin}
Define, for $t\geq0$ and $h>0$,
\begin{equation}\label{eq:A}
 A_s(t,h)=\frac{(t+h)^{s+2}-t^{s+2}-h^{s+2}}{h(s+1)(s+2)},\qquad
 b_\gamma(t,h)=(1-\gamma)L A_s(t,h).
\end{equation}
At $s=1$, $A_1(t,h)=t(t+h)/2$. The integral definition
\[
 A_s(t,h)=h^{-1}\int_0^t\int_0^h(x+u)^s\dd u\dd x
\]
is often more useful than the closed form for numerical evaluation at very small $t/h$.

\begin{lemma}[Counterfactual bridge]\label{lem:bridge}
For every law in \cref{ass:class} and every $t\in[0,dT]$,
\begin{equation}\label{eq:bridge}
 \left|\int_0^t g(x)\dd x-t\mu_h\right|\leq b_\gamma(t,h).
\end{equation}
Moreover $A_s(t,h)\leq t(t+h)^s$ and $A_s(t,h)\leq t(t^s+h^s)/(s+1)$. In particular,
\begin{equation}\label{eq:pseudobias}
 \theta_h=\frac{p-\gamma}{d\mu_h},\qquad
 |\theta_h-\theta|\leq \frac{LT}{c}(dT+h)^s\quad(d>0).
\end{equation}
\end{lemma}
The bridge uses the same log-coordinate density class as the structural model. It does not assume the observed upper density is a proportional copy of the lower density. Its allowance is an outer bound on possible integrals; we do not assert it is the sharp identified-set bound for every $s$.

\subsection{Two measures of difficulty}
\begin{corollary}[Population ambiguity order]\label{cor:ambiguity}
For sufficiently small $d>0$, the worst-case diameter of the population identified set is bounded above and below by positive constants times $d^s$, uniformly over known $\gamma\leq\bar\gamma$.
\end{corollary}
\begin{proof}
Choose $h=d$. Every compatible $\vartheta\in I(P)$ shares the same observed $p,q$ and hence the same $\theta_d=(p-\gamma)/(d\mu_d)$. Applying \eqref{eq:pseudobias} to its compatible density yields $|\vartheta-\theta_d|\leq(LT/c)(dT+d)^s$. The diameter is consequently at most twice this bound. The observationally equivalent pair in \cref{thm:equiv} supplies a compatible law whose identified set contains two values separated by $\kappa d^s$. The common background mixture does not change their equivalence.
\end{proof}
This population order separates counterfactual ambiguity from the sampling uncertainty introduced next. It is a worst-case class statement; a particular law can have a smaller identified set under additional restrictions.

The variance of a local-flat moment can change with the policy. At its population ratio center, define
\begin{equation}\label{eq:V}
 V_h=p-\gamma^2+\frac{(p-\gamma)^2}{q},\qquad
 v_h=\frac{V_h}{nd^2\mu_h^2}.
\end{equation}
This is the exact variance of a linearized influence function, not the finite-sample variance of a random ratio. Its inverse square root measures local stochastic sensitivity. Under vanishing $d,h$ and the local class,
\begin{equation}\label{eq:vapprox}
 v_h\asymp \frac{\gamma+d}{nd^2}+\frac1{nh},
\end{equation}
uniformly for $\theta\in\ThetaS$ and $\gamma\leq\bar\gamma$. With $h=d$, the scalar index
\begin{equation}\label{eq:concentration}
 \mathcal A_{n,d,\gamma}=\frac{nd^2}{\gamma+d}
\end{equation}
describes the stochastic component up to class constants. Large $\mathcal A$ means that a local-flat contrast can concentrate; bounded $\mathcal A$ means that uniform concentration is impossible even in a known-density submodel. We call the latter \emph{weak stochastic identification}. This label does not assert point identification under the unknown-density model.

The other component is the extrapolation allowance $d^s$. It remains at a fixed kink, regardless of sample size. A useful diagnostic therefore has two coordinates: a stochastic sensitivity or count-based uncertainty interval, and a smoothness allowance in elasticity units. A single first-stage $F$ statistic cannot encode both. Even when $\mathcal A$ diverges, Wald inference can fail because extrapolation bias exceeds its shrinking standard error.

When $\gamma=0$, $\mathcal A=nd$. When $\gamma$ is fixed and positive, $\mathcal A$ is proportional to $nd^2$. Between these regimes the crossover is governed by $\gamma/d$. The total observed atom mass, which includes $\gamma$, is not itself a valid behavioral strength diagnostic. This observation makes precise why ``small bunching mass'' and ``weak identification'' should not be treated as synonyms across observation mechanisms.

\section{Matching precision bounds for the full experiment}\label{sec:minimax}
Let $\mathcal R_{n,d,\gamma}$ be the minimax expected absolute loss,
\[
 \mathcal R_{n,d,\gamma}=
 \inf_{\widehat\theta}\sup_{\theta\in\ThetaS,\ f\in\Holder}
 \E_{\theta,f,d,\gamma}|\widehat\theta-\theta|,
\]
where the infimum is over all measurable estimators based on the complete sample and the known inputs $n,d,\gamma$. Constants in statements below may depend on the fixed class but never on $n,d,\gamma$. Define $\rho_{n,0,\gamma}=1$ and, for $d>0$, define $\rho_{n,d,\gamma}$ by \eqref{eq:intro-rate}.

\begin{theorem}[Minimax absolute-error risk]\label{thm:minimax}
Under \cref{ass:class}, there are $d_0>0$ and constants $0<K_1\leq K_2<\infty$ such that, for every $n\geq1$, every $0<d\leq d_0$, and every known $\gamma\in[0,\bar\gamma]$,
\begin{equation}\label{eq:minimax}
 K_1\rho_{n,d,\gamma}\leq\mathcal R_{n,d,\gamma}
 \leq\sup_{\theta,f}\E|\widetilde\theta-\theta|
 \leq K_2\rho_{n,d,\gamma},
\end{equation}
where $h=d$, $c_*=c(1-\bar\gamma)$, and
\begin{equation}\label{eq:regularized}
 \widetilde\theta=\clip_{\ThetaS}\left\{
 \frac{\widehat p-\gamma}{d\max(\widehat\mu_d,c_*/2)}\right\}.
\end{equation}
At $d=0$ the risk is bounded above and below by positive constants, and any fixed midpoint estimate has the stated constant order.
\end{theorem}
The projection and denominator floor are devices for a uniform risk bound, not recommended interpretations of the unregularized point estimate. They allow a negative net atom estimate, a zero bin count, and weak designs without a division error. The conventional estimator studied later is left unprojected so that its behavior is visible.

The lower bound has two independent parts. First, in a known density with a flat neighborhood and smooth tails, the squared Hellinger distance between complete observation laws satisfies
\begin{equation}\label{eq:Hbound}
 H^2(P_{\theta_1},P_{\theta_2})
 \leq K\frac{d^2(\theta_1-\theta_2)^2}{\gamma+d}.
\end{equation}
The atom supplies the denominator $\gamma+d$; the translated upper density contributes a term of order $d^2(\theta_1-\theta_2)^2$, which is also included. Product-measure affinity then bounds the distinguishability of two nearby elasticities. Thus the stochastic lower bound is valid even for an estimator exploiting every continuous observation. It would be incorrect to derive a lower bound only for atom counts and infer the same bound for the more informative full sample.

Second, \cref{thm:equiv} gives an indistinguishable pair separated by $\kappa d^s$. The pointwise triangle inequality under their common law forces positive estimation loss. Taking the maximum of these two lower bounds gives the sum in \eqref{eq:minimax}, up to fixed constants. Neither part relies on a particular density estimator.

For the upper bound, the net atom estimate has mean $p-\gamma=O(d)$ but variance $O((\gamma+d)/n)$. The bin density has expected absolute stochastic error $O((nd)^{-1/2})$. Its contribution to the ratio is dominated by the net atom's structural error rate. The bridge controls the remaining population discrepancy by $O(d^s)$. This proves that a separate nonparametric first-step rate need not become an additional leading obstacle in this experiment. The conclusion is derived from the weights in the moment, and should not be generalized to unrelated bunching estimators.

\begin{corollary}[Consistency and policy-size balance]\label{cor:balance}
For sequences obeying \cref{ass:class}, uniform consistent estimation over the class is possible when $d_n\to0$ and $\mathcal A_{n,d_n,\gamma_n}\to\infty$. If either $d_n$ stays at a sufficiently small fixed positive value or $\mathcal A$ stays bounded, uniform consistency is impossible. For a design choice that keeps the estimand and density class fixed, the two error terms balance at the unique positive solution of
\begin{equation}\label{eq:balance}
 n d^{2s+2}=\gamma+d.
\end{equation}
If $\gamma_n=O(n^{-1/(2s+1)})$, its order is $n^{-1/(2s+1)}$ and the attainable rate is $n^{-s/(2s+1)}$. If $\gamma_n/n^{-1/(2s+1)}\to\infty$, its order is $(\gamma_n/n)^{1/(2s+2)}$ and the rate is $(\gamma_n/n)^{s/(2s+2)}$.
\end{corollary}
The assertion about impossibility when $d$ does not vanish is a fixed-small-kink statement, not a claim for every arbitrary nonshrinking policy outside the local class. Equation \eqref{eq:balance} is a precision comparison, not an optimal-tax recommendation. A real design choice would also face welfare, implementation, and external-validity considerations. Nor does the result say that collecting data under several policy schedules is equivalent to one schedule: joint restrictions across schedules can add identification.

\begin{figure}[tb]
\centering\includegraphics[width=\textwidth]{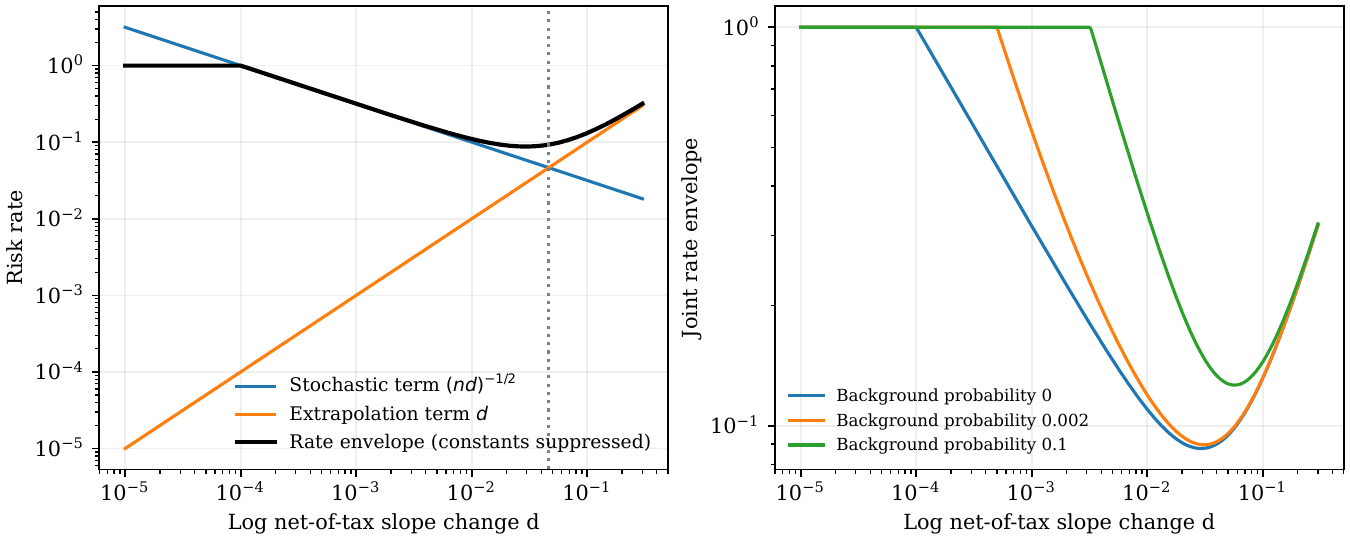}
\caption{Risk-rate expressions for $s=1$ and $n=10{,}000$. Left: stochastic and extrapolation components for $\gamma=0$. Right: joint envelopes for three known background probabilities. Constants are suppressed; these are analytical expressions, not simulated or estimated risks.}
\label{fig:rate}
\end{figure}
\begin{table}[tb]
\centering\small
\begin{tabular}{lll}
\toprule
Observation regime & Stochastic error order & Concentration order\\
\midrule
$\gamma=O(d)$ & $(nd)^{-1/2}$ & $nd$\\
$\gamma/d\to\infty$ & $\sqrt{\gamma}/(\sqrt n d)$ & $nd^2/\gamma$\\
$\gamma>0$ fixed & $(nd^2)^{-1/2}$ & $nd^2$\\
\bottomrule
\end{tabular}
\caption{Background-mass regimes in the same complete-observation theorem. Every row also carries the population ambiguity order $d^s$. A transition in sampling information does not eliminate counterfactual uncertainty.}
\label{tab:regimes}
\end{table}

The kink size behaves like an identification window. A larger window contains more movers but also a larger unobserved stretch of the counterfactual distribution. This structural extrapolation cost persists even when both tails are observed without error. The result is particular to the unknown-density class: with a known counterfactual distribution the $d^s$ term disappears and a larger incentive can improve precision over the same local range.

\section{Conventional estimation and distributional regimes}\label{sec:conventional}
Consider the local-flat estimator
\begin{equation}\label{eq:ratio}
 \widehat\theta_h=\frac{\widehat p-\gamma}{d\widehat\mu_h},\qquad
 \widehat v_h=\frac{\widehat p-\gamma^2+(\widehat p-\gamma)^2/\widehat q}
 {nd^2\widehat\mu_h^2}.
\end{equation}
It is defined only for $d>0$ and $\widehat q>0$. A reported software failure in an empty bin is not silently replaced with a successful estimate. In the simulations an undefined conventional interval is recorded as unavailable. The proposed confidence set is still defined.

\begin{theorem}[Normal and rare-count limits]\label{thm:limits}
Under \cref{ass:class}, consider a triangular sequence with $d_n,h_n>0$.
\begin{enumerate}[label=(\roman*)]
\item If $np_n\to\infty$ and $nq_n\to\infty$, then
\[
 (\widehat\theta_h-\theta_h)/\sqrt{v_h}\ \Rightarrow\ N(0,1),\qquad
 \widehat v_h/v_h\ \to_{\Prob}\ 1.
\]
These conclusions hold uniformly over classes in which both expected counts diverge uniformly. Structural centering at $\theta$ also holds if $|\theta_h-\theta|/\sqrt{v_h}\to0$ uniformly. The sufficient class condition is $(d^s+h^s)/\sqrt{v_h}\to0$.
\item Suppose $\gamma_n=0$, $d_n\to0$, $h_n\to0$, $nd_n\to\tau\in(0,\infty)$, $nh_n\to\infty$, $\theta_n\to\theta>0$, and $f_n(0)\to f_*>0$. Then
\[
 N_0\Rightarrow N\sim\Pois(\tau\theta f_*),\qquad
 \widehat\theta_h\Rightarrow N/(\tau f_*).
\]
If instead $nd_n\to0$ while the denominator conditions remain, $\widehat\theta_h\to_{\Prob}0$, and it is inconsistent for every positive limiting $\theta$.
\item Under the conditions of (ii), replace $\gamma_n=0$ by $n\gamma_n\to\zeta\geq0$. Then the ratio limit is $(N-\zeta)/(\tau f_*)$ with $N\sim\Pois(\zeta+\tau\theta f_*)$.
\end{enumerate}
For precision, let $\mathcal Q_n$ be any subclasses with $\inf_{P\in\mathcal Q_n}np_P\to\infty$ and $\inf_{P\in\mathcal Q_n}nq_P\to\infty$. The uniform assertion in (i) means
\[
 \sup_{P\in\mathcal Q_n}\sup_z\left|\Prob_P\{(\widehat\theta_h-\theta_h)/\sqrt{v_h}\leq z\}-\Phi(z)\right|\to0,
\]
and $\sup_{P\in\mathcal Q_n}\Prob_P(|\widehat v_h/v_h-1|>\delta)\to0$ for every fixed $\delta>0$. Undefined ratios may be assigned an arbitrary value on the empty-bin event, whose probability vanishes uniformly. Structural centering additionally requires $\sup_{P\in\mathcal Q_n}|\theta_h-\theta|/\sqrt{v_h}\to0$.
\end{theorem}
The normal conclusion in (i) is about an unprojected ratio and its population pseudo-center. It does not require consistency for the structural parameter. In particular, a deliberately noisy first step may yield a normal approximation with a large variance. At a parameter-space boundary, projection changes the distribution, so the same normal statement cannot simply be attached to \eqref{eq:regularized}.

With $\gamma=0$ and $h=d\to0$, $v_h$ has order $(nd)^{-1}$. Thus $nd\to\infty$ gives a normal approximation around $\theta_h$, but ignoring the worst-case extrapolation discrepancy requires $nd^{2s+1}\to0$. At the minimax balance the discrepancy and standard error have the same order. In the ambiguity-dominated region, a small conventional standard error does not imply accurate inference on $\theta$.

For a fast atom-dominated variance, one can take $h/d\to\infty$, $nd\to\infty$, and $\sqrt{nd}(d^s+h^s)\to0$. Such a bin exists if $nd^{2s+1}\to0$. This condition is \emph{not} necessary for every centered normal procedure: for $s=1$, $d=n^{-1/5}$ and $h=n^{-9/10}$ give density-dominated standard deviation of order $n^{-1/20}$ while the bias is $O(n^{-1/5})$. The estimator is centered normally at a deliberately slow rate even though $nd^3$ diverges. This distinction prevents a bandwidth condition for a fast procedure from being mislabeled a universal impossibility result.

When $\gamma$ is fixed and positive, $d=\kappa/\sqrt n$, $\kappa>0$, $nh\to\infty$, $h\to0$, $\theta_n\to\theta$, and $f_n(0)\to f_*>0$, the count-based ratio has the nonconcentrating normal limit
\begin{equation}\label{eq:bg-normal}
 \widehat\theta_h\Rightarrow \theta+
 \frac{\sqrt{\gamma(1-\gamma)}}{\kappa(1-\gamma)f_*}Z,
 \qquad Z\sim N(0,1),
\end{equation}
provided $d/h$ stays bounded, which makes the density term negligible here. This is a limit for the stated estimator, not a characterization of the complete Gaussian experiment. In a known-density submodel, the translated continuous upper observations also carry information of order $nd^2$.

\begin{theorem}[Coverage failure of a plug-in Wald interval]\label{thm:wald}
In \cref{thm:limits}(ii), let the conventional interval be
\[
 W_n=[\widehat\theta_h-z_{1-\alpha/2}\sqrt{\widehat v_h},
 \widehat\theta_h+z_{1-\alpha/2}\sqrt{\widehat v_h}],
\]
with a zero estimated variance interpreted as a singleton and an unavailable interval counted as noncoverage. Put $\lambda=\tau\theta f_*$. Then
\begin{equation}\label{eq:wald-bound}
 \limsup_n\Prob(\theta_n\in W_n)\leq1-e^{-\lambda}.
\end{equation}
For $\lambda<-\log\alpha$ the upper bound is below $1-\alpha$. Consequently this interval cannot maintain uniform nominal coverage over a class containing these sequences.
\end{theorem}
The mechanism is exact: when $N_0=0$ and $N_->0$, the point estimate and the estimated variance both equal zero. A positive elasticity is excluded. The result concerns the specified conventional formula; it does not say every method using a normal approximation fails in every weak design. Nor does it establish the actual coverage solely from the upper bound.

An Anderson--Rubin-style principle suggests evaluating a candidate moment rather than estimating its zero and dividing by its derivative. That principle is useful but insufficient by itself. A Gaussian critical value for a moment with a bounded expected number of atoms still uses the wrong limit law. The next section uses exact probability bounds instead.

\section{Feasible, honest inference with an estimated counterfactual}\label{sec:honest}
\subsection{Exact test inversion}
For $K\sim\operatorname{Bin}(n,u)$, let $[\ell_\beta(K),u_\beta(K)]$ be the central Clopper--Pearson interval with error probability $\beta$. It inverts the two binomial tails, using endpoint zero when $K=0$ and endpoint one when $K=n$. It has coverage at least $1-\beta$ for every success probability. Choose positive, prespecified $\alpha_p,\alpha_q$ with $\alpha_p+\alpha_q=\alpha<1/2$ and let
\[
 [p_L,p_U]=[\ell_{\alpha_p}(N_0),u_{\alpha_p}(N_0)],\qquad
 [q_L,q_U]=[\ell_{\alpha_q}(N_-),u_{\alpha_q}(N_-)].
\]
The counts are dependent, but their binomial marginal bounds are valid. Independence of the two intervals is neither assumed nor needed.

Define
\begin{equation}\label{eq:CS}
 C_n=\left\{\vartheta\in\ThetaS:
 \begin{array}{l}
 p_U\geq\gamma,\\
 p_L-\gamma\leq d\vartheta q_U/h+b_\gamma(d\vartheta,h),\\
 p_U-\gamma\geq d\vartheta q_L/h-b_\gamma(d\vartheta,h)
 \end{array}\right\}.
\end{equation}
The feasibility constraint $p_U\geq\gamma$ uses the known fact $p\geq\gamma$. It removes spurious acceptance from a negative lower approximation to the behavioral mass and preserves coverage on the simultaneous probability event.
For a candidate value $\vartheta$, reject when it is not in $C_n$. The procedure evaluates inequalities in the moment rather than using $1/\widehat J$. It is defined at every count realization, including $N_-=0$. A larger valid radius for the scaled density $g$ may replace $(1-\gamma)L$, yielding a conservative version. This is the version used in the simulations, where the common radius $0.5$ bounds every scaled normal density considered.

\begin{theorem}[Uniform finite-sample coverage]\label{thm:coverage}
For every $n\geq1$, every admissible $d\geq0$ and $h>0$, every known $\gamma\in[0,\bar\gamma]$, and every compatible population law $P$ under \cref{ass:class},
\begin{equation}\label{eq:honesty}
 \Prob_P\{I(P)\subseteq C_n\}\geq1-\alpha.
\end{equation}
In particular, the candidate test has size at most $\alpha$ and
\[
 \inf_{\theta,f,d,\gamma}\Prob_{\theta,f,d,\gamma}(\theta\in C_n)\geq1-\alpha.
\]
The infimum is finite-sample and includes strong, weak, and zero incentives. Prespecified inputs are required; selection from the same data is not covered by this theorem.
\end{theorem}
On the event that the two probability intervals contain their population probabilities, \cref{lem:bridge} implies acceptance of every observationally compatible parameter. A union bound gives the probability of that event. No exchange of limits, Gaussian approximation, or first-step negligibility is used. Covering the entire identified set is a stronger assertion than covering one true representative, and it follows here from the same simultaneous event.

At $d=0$, if $\gamma=0$ then $N_0=0$ almost surely and $C_n=\ThetaS$. With $\gamma>0$, the set is $\ThetaS$ or empty according to whether $\gamma\in[p_L,p_U]$. A rare rejection of a completely uninformative design is possible because the test also checks its known background probability. This distinction should be visible in software and in interpretation. A full parameter-space confidence set reflects limited information; an empty set can signal sampling rejection or incompatibility with the restrictions.

\subsection{Optimal order of expected diameter}
Let $\mathcal H_{n,d,\gamma}(\alpha)$ contain all measurable confidence sets that cover every true parameter with probability at least $1-\alpha$ over the stated model. Set $\diam(\varnothing)=0$. The lower bound below needs only pointwise honesty, although \eqref{eq:CS} also covers the population identified set.

\begin{theorem}[Matching honest-diameter bound]\label{thm:diameter}
Under \cref{ass:class}, for fixed $\alpha_p,\alpha_q>0$ and $\alpha<1/2$, there are positive constants $K_3,K_4$ and $d_0>0$ such that, for all $n\geq1$, $0<d\leq d_0$, and known $\gamma\leq\bar\gamma$,
\begin{equation}\label{eq:diameter}
 K_3\rho_{n,d,\gamma}\leq
 \inf_{C\in\mathcal H_{n,d,\gamma}(\alpha)}\sup_{\theta,f}\E\diam(C)
 \leq\sup_{\theta,f}\E\diam(C_n)
 \leq K_4\rho_{n,d,\gamma},
\end{equation}
where $C_n$ uses $h=d$. The same constant-order comparison holds at $d=0$.
\end{theorem}
The theorem matches diameter orders; it does not minimize the length constants associated with an error allocation or exact binomial intervals. These intervals are often conservative. The expectation bound is unconditional, including rejection events. It does not follow merely from coverage plus a high-probability width bound, which could leave a nonshrinking contribution from the coverage error.

For two accepted values $\vartheta_1<\vartheta_2$, combining the lower inequality at $\vartheta_2$ with the upper inequality at $\vartheta_1$ gives
\begin{equation}\label{eq:diamdet}
 d(q_L/h)(\vartheta_2-\vartheta_1)
 \leq(p_U-p_L)+dT(q_U-q_L)/h+2LdT(dT+h)^s.
\end{equation}
The bound is valid even for disconnected sets and does not require acceptance of the true parameter. Exact binomial interval widths, together with an exponentially small probability that $q_L/h$ falls below $c_*/2$, give the expected upper bound. Full-law testing and observational equivalence give the lower bound.

\subsection{Computing the set}
At $s=1$, the allowance is quadratic in $t=d\vartheta$. The upper envelope $tq_U/h+Lt(t+h)/2$ is increasing, while the lower envelope $tq_L/h-Lt(t+h)/2$ is concave. Solving their inequalities produces at most two closed intervals on a compact parameter space. The implementation returns all components. Their total length and their span are recorded separately. It never discards a far component or substitutes a convex hull for the computed set. Roots are evaluated using rationalized formulas to avoid subtracting nearly equal positive numbers.

For $s\in(0,1)$, the theoretical set is defined by \eqref{eq:CS}. A reliable numerical implementation would bracket all roots or use verified interval bounds. The supplied analytic component solver is for $s=1$; the general-smoothness theorems do not imply that the code implements arbitrary fractional $s$. An unbounded parameter space would also require a separate diameter formulation. We report all components within the declared $[0.1,2]$ simulation space.

\subsection{An independently estimated background atom}
Suppose a genuine, policy-neutral control design produces $m$ independent observations whose zero probability is the same $\gamma$. A third exact binomial interval $[\gamma_L,\gamma_U]$ can be included. Allocate three positive error probabilities summing to $\alpha$, form the joint acceptance set over $(\vartheta,\widetilde\gamma)$ by replacing $\gamma$ in \eqref{eq:CS} with $\widetilde\gamma\in[\gamma_L,\gamma_U]$, and project onto $\vartheta$.

\begin{proposition}[Control-sample coverage]\label{prop:control}
If the stated control model holds, the projected set covers the true elasticity with probability at least $1-\alpha$, uniformly over admissible treatment and control laws and every $n,m\geq1$.
\end{proposition}
The proof is the same simultaneous-event argument, now with three marginal intervals. This proposition is a coverage result. We do not prove a minimax rate in $(n,m,d,\gamma)$, nor assert that $\gamma$ can be identified from treatment data alone. Earlier bunching work already studies background bunchers and uses auxiliary information to estimate their mass \citep{goff2021}; introducing or subtracting such a mass is not claimed as new here.

\subsection{Semiparametric interpretation}
The moment in \eqref{eq:exact-moment} contains an infinite-dimensional nuisance $g$. The bridge replaces estimation of the entire unobserved path by a controlled inequality involving its observable bin mean. This changes the inference target from an assumed point-identified inverse map to a confidence set for compatible structural parameters. Cross-fitting is unnecessary for the two marginal count bounds. Orthogonalization would not eliminate the deterministic uncertainty over an interval where the density is unobserved.

More generally, a known monotone relocation map $t=G_d(\vartheta)$ can replace $d\vartheta$ in the bridge and test. Finite-sample coverage follows immediately under the same support and smoothness conditions. A diameter or minimax result for that model would additionally require quantitative bounds on $G_d$, its inverse, and the complete observation deformation. We therefore retain the canonical map for all precision claims.

\section{Structural Monte Carlo evidence}\label{sec:mc}
\subsection{Design and implementation}
The baseline counterfactual is $X\sim N(\mu,1)$ with $\mu\in\{-0.75,0,0.75\}$. Its positive and negative local slopes, together with curvature, vary across these locations. Income choices are generated by \eqref{eq:choice}. This can be implemented by drawing $X$, setting productivity from \eqref{eq:transformation}, and solving the already derived strictly concave choice problem. For the reported procedures, only the atom and unaffected-bin counts matter. We therefore draw their exact multinomial probabilities from the same structural law, avoiding needless storage of individual observations. A separate test generates individual choices and verifies these probabilities. This shortcut changes computation, not the data-generating model.

Sample sizes are $1{,}000$, $5{,}000$, and $20{,}000$. The fixed-kink design uses $d=0.15$. The three shrinking designs use $d=25/(n\theta f(0))$, $2/(n\theta f(0))$, and $0.2/(n\theta f(0))$, respectively. These target approximately the stated expected numbers of behavioral bunchers; actual probabilities are evaluated exactly, without a small-kink approximation. A zero-kink design is included. We take $\theta=0.6$, $h\in\{0.5,1,2\}n^{-1/3}$, and parameter space $[0.1,2]$. The radius $L=0.5$ exceeds the global Lipschitz constant $e^{-1/2}/\sqrt{2\pi}$ of every normal density in these designs. The same radius also bounds its scaling by $1-\gamma$.

There are \MCDesigns{} baseline designs, each with \MCRepsPerDesign{} repetitions, for \MCReplications{} executed replications. A second experiment has 45 designs and 225,000 replications, varying known background probabilities $0$, $0.002$, and $0.1$ at the three sample sizes. Its shrinking kink solves $nd^2/(\gamma+d)\in\{25,2,0.2\}$; actual behavioral and total atom counts are both retained. Each design has its own fixed seed. The code reports unavailable Wald intervals rather than imputing successes. Under zero incentives the ratio is undefined, so its recorded noncoverage is accompanied by an unavailable indicator and is not interpreted as a numerical Wald size estimate.

We compare the unprojected local-flat ratio, its correctly covaried Wald interval, and the honest set. All confidence procedures use nominal level $0.95$. The two exact intervals split error equally. An additional infeasible Gaussian moment comparison uses the true null-moment variance, to isolate distributional and approximation errors; it is not proposed as an implementable competitor. The raw files contain bias, root mean squared error (RMSE), median absolute error, coverage, set length, diameter, zero-count frequency, disconnected and empty-set frequency, and conventional availability. Rejection probability is one minus coverage at the true null.

\subsection{Coverage and length}
Bias, RMSE, median absolute error, and Wald length are computed conditional on an available conventional estimator; its unavailable fraction is reported separately. Coverage includes all replications, counting an unavailable Wald interval as noncoverage. The nonzero baseline designs in this run have no unavailable estimates. These conventions prevent a favorable conditional coverage report from concealing failures.
\begin{table}[tb]
\centering\small
\begin{tabular}{rrlrrrrr}
\toprule
$n$ & $E[N_0]$ & Regime & Bias & RMSE & Wald & Honest & Length \\
\midrule
1,000 & 35.86 & Strong & 0.015 & 0.149 & 0.942 & 1.000 & 1.453 \\
1,000 & 24.98 & Moderate & 0.014 & 0.159 & 0.944 & 1.000 & 1.457 \\
1,000 & 2.00 & Weak & 0.022 & 0.456 & 0.862 & 0.999 & 1.867 \\
1,000 & 0.20 & Very weak & -0.001 & 1.372 & 0.177 & 0.999 & 1.895 \\
5,000 & 179.28 & Strong & 0.004 & 0.073 & 0.949 & 1.000 & 0.690 \\
5,000 & 25.00 & Moderate & 0.007 & 0.136 & 0.938 & 0.999 & 0.960 \\
5,000 & 2.00 & Weak & 0.019 & 0.439 & 0.872 & 0.999 & 1.836 \\
5,000 & 0.20 & Very weak & -0.008 & 1.351 & 0.177 & 0.999 & 1.894 \\
20,000 & 717.13 & Strong & 0.001 & 0.041 & 0.951 & 1.000 & 0.389 \\
20,000 & 25.00 & Moderate & 0.004 & 0.124 & 0.939 & 0.997 & 0.786 \\
20,000 & 2.00 & Weak & -0.010 & 0.425 & 0.858 & 0.995 & 1.806 \\
20,000 & 0.20 & Very weak & 0.041 & 1.398 & 0.191 & 0.998 & 1.892 \\
\bottomrule
\end{tabular}

\caption{Baseline results with $\mu=0$, $h=n^{-1/3}$, and 5,000 replications per row. Wald and honest columns are coverage probabilities. Length is mean total honest-set length in the declared parameter space. ``Strong'' labels the fixed-kink count design, not point identification under an unknown H\"older density. All entries are generated from the raw simulation summary.}
\label{tab:baseline}
\end{table}

Across baseline weak-count designs, Wald coverage ranges from \WeakWaldMin{} to \WeakWaldMax{}. In the very-weak designs it ranges from \VeryWeakWaldMin{} to \VeryWeakWaldMax{}. The zero-count event explains much of the latter failure. The minimum honest coverage across the baseline designs is \HonestCoverageMin{}. This is protection with a substantial conservativeness cost, not evidence of an exactly calibrated $0.95$ rejection rate. At 5,000 repetitions the Monte Carlo standard error at a coverage probability of $0.95$ is approximately \NominalCoverageMCSE{}; the maximum possible standard error for a reported probability is \MaxCoverageMCSE{}.

\begin{figure}[tb]
\centering\includegraphics[width=\textwidth]{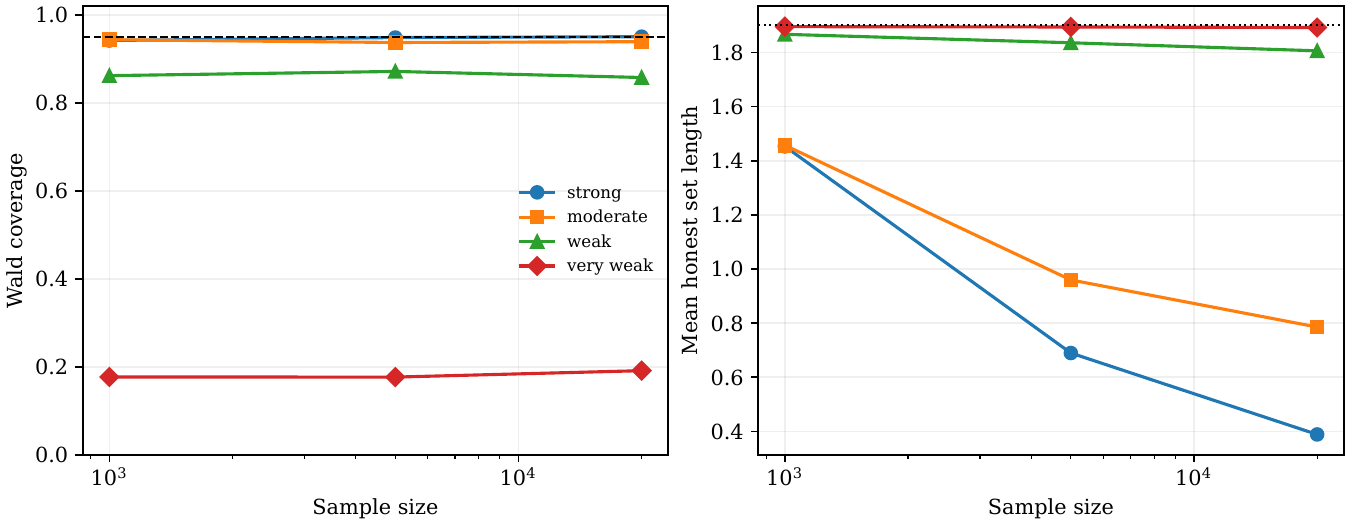}
\caption{Executed baseline simulations with centered normal counterfactual density and $h=n^{-1/3}$. The left panel reports conventional Wald coverage; the right panel reports mean total honest-set length. The dotted length line is the full parameter-space length. Coverage protection can entail nearly uninformative sets in weak designs.}
\label{fig:coverage}
\end{figure}

The fixed-kink design separates stochastic strength from population ambiguity. Counts become more precise as $n$ increases, but the approximation of an unobserved interval by a local-flat density need not improve. With noncentered counterfactual densities, its structural bias can become visible relative to a shrinking standard error. The honest procedure includes that uncertainty. In shrinking designs, a larger bin reduces density noise but increases the permitted extrapolation discrepancy. These are substantive tuning costs rather than failures to optimize an arbitrary numerical criterion.

\begin{figure}[tb]
\centering\includegraphics[width=\textwidth]{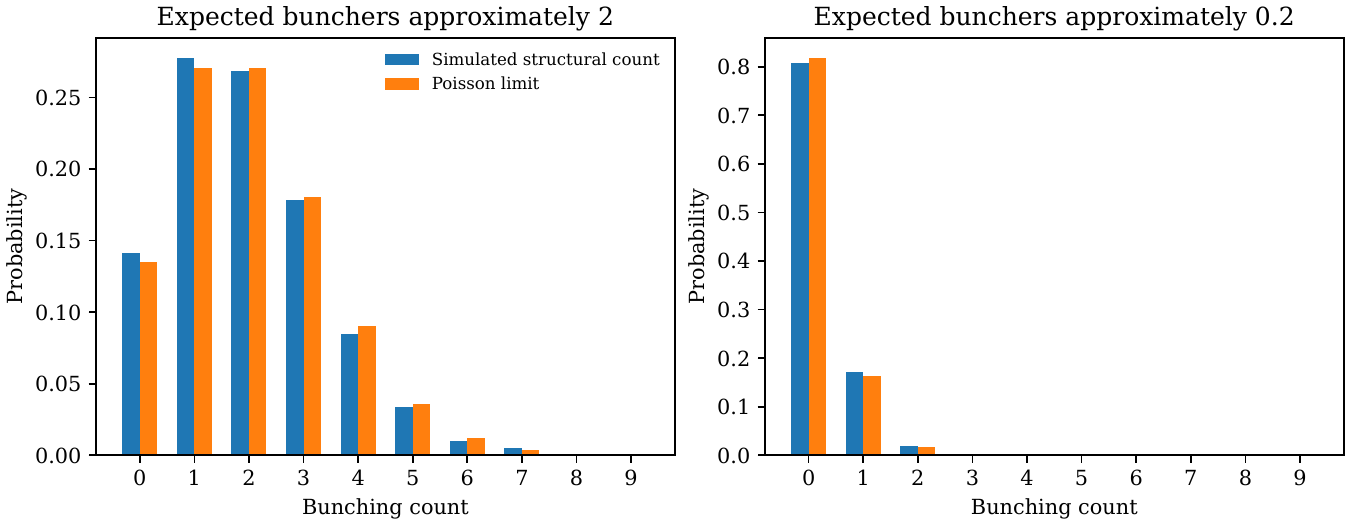}
\caption{Empirical frequencies from the executed structural Monte Carlo at $n=20{,}000$, compared with the Poisson probabilities for the two target expected-count designs. The approximation concerns atom counts. The full likelihood-ratio statement is proved separately in \cref{thm:poissonlr}.}
\label{fig:rare}
\end{figure}

The background experiment illustrates why a large recorded atom need not imply strong behavioral information. Subtracting a known background probability leaves sampling noise in the total count. Holding $nd^2/(\gamma+d)$ fixed therefore differs from holding the number of behavioral bunchers fixed. The complete output is provided for all background designs, including the many-atom Gaussian regime. These simulations verify formulas and finite-sample behavior; they do not establish a universal calibrated cutoff for the strength diagnostic.

\subsection{Background evidence}
Table \ref{tab:background} holds the concentration target fixed across background probabilities. It reports both behavioral and total atom counts, keeping their distinct roles visible. Many background observations improve the count's normal approximation while leaving weak structural precision; this explains why a conventional normal interval can have substantially better coverage in these rows without a narrow honest set.
\begin{table}[tb]
\centering\footnotesize
\begin{tabular}{rrrrrrrr}
\toprule
$\gamma$ & $\mathcal A$ & Total & Behavioral & Wald & Honest & Length & Diameter \\
\midrule
0.000 & 25.0 & 5.98 & 5.98 & 0.919 & 0.997 & 1.550 & 1.550 \\
0.000 & 2.0 & 0.48 & 0.48 & 0.392 & 0.999 & 1.890 & 1.890 \\
0.000 & 0.2 & 0.05 & 0.05 & 0.046 & 0.999 & 1.898 & 1.898 \\
0.002 & 25.0 & 17.80 & 7.80 & 0.935 & 0.995 & 1.641 & 1.641 \\
0.002 & 2.0 & 11.33 & 1.33 & 0.926 & 0.990 & 1.866 & 1.866 \\
0.002 & 0.2 & 10.36 & 0.36 & 0.942 & 0.984 & 1.876 & 1.876 \\
0.100 & 25.0 & 526.93 & 26.93 & 0.956 & 0.989 & 1.744 & 1.744 \\
0.100 & 2.0 & 507.03 & 7.03 & 0.947 & 0.978 & 1.848 & 1.848 \\
0.100 & 0.2 & 502.18 & 2.18 & 0.956 & 0.979 & 1.859 & 1.859 \\
\bottomrule
\end{tabular}

\caption{Executed background experiment at $n=5{,}000$, $h=n^{-1/3}$, and 5,000 replications per row. Total and Behavioral are expected recorded atom and net behavioral counts; Wald and Honest are coverage. Length is total component length and Diameter is its span. Rows are ordered by background probability and concentration target. Large total atom counts can coexist with weak structural precision.}
\label{tab:background}
\end{table}

\subsection{An exactly equivalent structural pair}
An additional experiment directly illustrates population ambiguity. Set $f_1$ uniform on $[-2,2]$ and $\theta_1=0.6$. For each chosen $d$, let $\theta_2=\theta_1+0.1d$ and modify $f_1$ on $[0,d\theta_2]$ by the quadratic bump in the equivalence proof, translating its positive tail by $0.1d^2$. The complete recorded distribution is identical under these two optimizing populations. Both densities satisfy the stated local Lipschitz radius $0.5$, with a positive lower bound $0.1$ near the threshold. Their differing elasticity cannot be distinguished by observing more of that same distribution.

Sample sizes are $500$, $2{,}000$, $8{,}000$, and $32{,}000$, with known backgrounds $0$, $0.002$, and $0.1$. We choose $d$ from $nd^4=\gamma+d$, take $h=d$, and use the projected estimator with the common floor based on $c=0.1$ and $\bar\gamma=0.1$. There are \PairDesigns{} designs and \PairReplications{} executed replications. The recorded error averages absolute loss across the two compatible elasticities, and the set is evaluated for simultaneous inclusion of both. Its minimum observed joint coverage is \PairJointCoverageMin{}. Together the three experiments contain \AllReplications{} executed replications.

\begin{figure}[tb]
\centering\includegraphics[width=\textwidth]{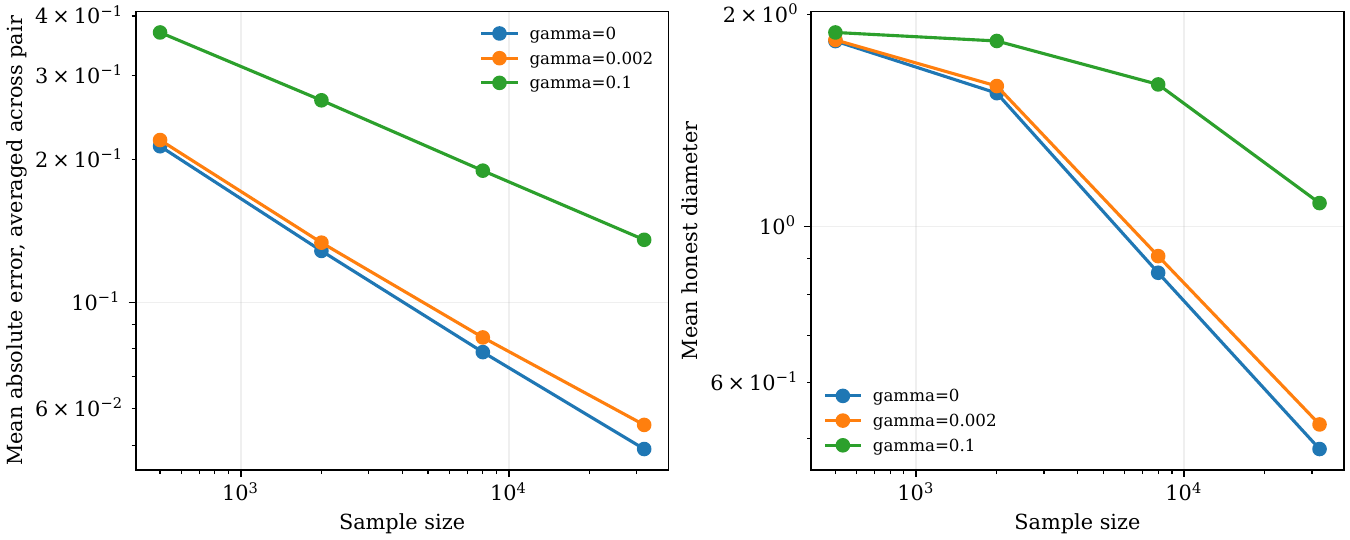}
\caption{Executed simulations from exactly equivalent optimizing populations. Left: projected estimator error averaged over the pair. Right: mean honest-set diameter. Each point uses 5,000 replications. The kink changes with sample size according to the precision balance; this experiment illustrates the theorem and does not estimate the supremum risk over the density class.}
\label{fig:equivalence}
\end{figure}
For every realized estimate, its two absolute errors average at least half the parameter separation, by the triangle inequality. This holds before averaging simulations and would still hold for an estimator using the complete continuous sample. The experiment therefore distinguishes intrinsic ambiguity from a failure of histogram estimation. It does not identify least-favorable constants or prove the lower bound numerically.

\subsection{Scope of the numerical evidence}
The normal data-generating distributions need not attain the least-favorable extrapolation ambiguity in the minimax theorem. They are smooth and globally identifiable if their parametric family is assumed known. The procedures intentionally use only a stated local smoothness restriction. The theorem's lower bound is therefore established by the equivalence construction, not by failure to estimate a normal distribution in these simulations. No claim of finite-sample minimax constants is based on the tables.

The simulations do not include data-dependent selection of the smoothness radius, arbitrary income rounding, estimated measurement errors, multiple thresholds, or heterogeneous elasticities. They also do not treat a synthetic example as an empirical application. The exact component solver is checked against direct inequality evaluation, including cases with two components; component counts in the raw output are retained even when a particular design produces none.

\section{Practical implications and conclusion}
If $\gamma$ is calibrated rather than literally known, state the policy-neutral population used for calibration, explain why its threshold probability equals the treated background probability, and quantify its sampling uncertainty. Finite calibration error calls for the control-sample projection; the known-$\gamma$ diameter theorem cannot then be used unchanged. Neither an exact probability interval nor a good outside-density fit establishes this cross-population identity.
A bunching analysis using these results must specify what is observed at the threshold. Exact atom counts, a background mixture with an independently justified calibration, and broadly rounded income are different experiments. A count within a nonzero bin is not automatically a behavioral atom. In the latter case, its probability includes continuous background observations, and a new deformation and bias analysis is required. The present theorem does not turn an arbitrary histogram spike into point-identified elasticity information.

For an exact-atom analysis, report the policy size $d$, the total atom count, the estimated counterfactual bin probability, the declared parameter range, and the smoothness allowance. If a background mass is included, state how it is known or provide the control-sample confidence interval. Reporting an elastic point estimate without its net-atom count and density uncertainty conceals the source of precision. The population analogue of the diagnostic is $\mathcal A_{n,d,\gamma}$ together with $d^s$; a feasible analysis can report probability intervals and the implied moment allowance instead of attaching a universal strong/weak cutoff to a point estimate.

Prespecify the unaffected bin and examine sensitivity to several justified radii or widths. Separate sensitivity reports are descriptive unless a simultaneous selection correction is proved. If the goal is a single honest set after inspecting several widths, a conservative finite collection of bins can receive separate error allocations before inspection, and their confidence sets can be intersected. A union bound then maintains coverage. This construction can be costly; an optimized adaptive procedure is outside the present results. Smoothness restrictions cannot be validated inside the unobserved relocation interval merely by obtaining a good fit outside it.

The finite-sample method is straightforward to implement at $s=1$, but its relative simplicity should not obscure the identifying restriction. Exact sampling control protects against weak information; it does not protect against a false counterfactual class or a misdescribed observation mechanism. If the set is wide, the substantive conclusion is limited information under that class. If it is disconnected, report the components. If it is empty, disclose the rejection and check the maintained model; do not select a convenient nonempty component or change a tuning parameter until a preferred elasticity appears.

This paper establishes a matching precision bound for a local bunching experiment with an unknown counterfactual density. Sampling uncertainty has two observation-dependent scales, while population ambiguity comes from the unobserved relocation interval. Both are relevant to a credible confidence set. The main mathematical contribution is their combination in full-observation lower bounds and matching feasible upper bounds, including background-atom transitions. The exact inversion is an implementation of classical probability tools tailored to this structural geometry.

Several extensions remain substantive research problems. Unknown background mass introduces a second sample-size dimension. Diffuse frictions and general coarsening remove the exact behavioral atom. Higher smoothness orders require estimating or canceling lower-order density terms, because the zeroth-order bridge used here directly covers only $0<s\leq1$. Observed policy variation across several kinks can identify information that is absent from one schedule. None of these settings inherits the complete minimax theorem merely from having a similar-looking moment condition.

\subsection*{Computational disclosure}
All numerical experiments reported here were executed locally using fixed seeds. The replication files generate the tables, figures, and numerical statements from saved outputs. No private tax data are used. Artificial intelligence (AI) assistance through OpenAI Codex was used for literature discovery, model development, proof construction and checking, code, and drafting; no AI system is listed as an author. Such assistance does not substitute for the author's responsibility for the manuscript or for external scholarly review.

\appendix
\section{Proofs}\label{app:proofs}
Throughout the appendix, constants are finite and depend only on the fixed class and fixed confidence-error allocations. They can change between displays. All probability calculations use the full mixed observation law unless a marginal count is explicitly stated.

\subsection{The bridge and the observational-equivalence construction}\label{app:equiv}
\begin{proof}[Proof of \cref{lem:bridge}]
By Fubini on a bounded rectangle,
\[
 \int_0^t g(x)\dd x-t\mu_h
 =h^{-1}\int_0^t\int_0^h[g(x)-g(-u)]\dd u\dd x.
\]
The integrand's absolute value is at most $(1-\gamma)L(x+u)^s$. Integrating twice gives \eqref{eq:A} and \eqref{eq:bridge}. Since $x+u\leq t+h$, the first upper bound follows. For $0<s\leq1$, $(x+u)^s\leq x^s+u^s$; integrating gives the second. At $t=d\theta$, divide by $d\mu_h\geq dc(1-\gamma)$ to obtain \eqref{eq:pseudobias}. Every integration occurs within the declared density neighborhood. No extension of the restriction outside that neighborhood is used.
\end{proof}

\begin{proof}[Proof of \cref{thm:equiv}]
Let $v=1/4$, $\theta_*=(a+T)/2$, and choose a density $f_1$ equal to $v$ on $[-1,1]$. A smooth global choice is
\begin{equation}\label{eq:plateau}
 f_1(x)=v\exp\left\{-\frac{((|x|-1)_+)^2}{2\sigma^2}\right\},
 \qquad\sigma=\sqrt{2/\pi}.
\end{equation}
Its integral is $2v+v\sigma\sqrt{2\pi}=1$. Fix $\kappa>0$ so small that $6v\kappa/\theta_*^{s+1}\leq L$. For $d>0$ put
\[
 \delta=\kappa d^s,\quad \theta_1=\theta_*,\quad\theta_2=\theta_*+\delta,
 \quad t_j=d\theta_j,\quad A=\frac{3v\delta}{2(\theta_*+\delta)}.
\]
Choose $d_0$ small enough that $\theta_2<T$, $t_2<\varepsilon$, $d_0\leq\varepsilon$, and $A\leq v-c$. Let $b(u)=4u(1-u)$ on $[0,1]$, extended by zero elsewhere, and define
\begin{equation}\label{eq:f2}
 f_2(x)=\begin{cases}
 f_1(x),&x<0,\\v-A b(x/t_2),&0\leq x\leq t_2,\\
 f_1(x-d\delta),&x>t_2.
 \end{cases}
\end{equation}
The boundary values at zero and $t_2$ are $v$. Since $\int_0^1b(u)\dd u=2/3$,
\[
 \int_0^{t_2}f_2(x)\dd x=vt_2-2At_2/3=vd\theta_*.
\]
The upper integral is $\int_{t_1}^{\infty}f_1(x)\dd x$. The lower integral is unchanged. Their sum with the displayed middle integral is one. The construction is nonnegative globally; on the local neighborhood its values lie in $[c,v]\subset[c,C]$.

The zero-extended $b$ has range at most one and Lipschitz constant at most four. For all real $u,w$, $|b(u)-b(w)|\leq\min\{4|u-w|,1\}\leq4|u-w|^s$. Inside $[-\varepsilon,\varepsilon]$, the continuation in \eqref{eq:f2} is the constant $v$ outside the bump. Hence its H\"older constant is at most
\[
 4A/t_2^s=6v\kappa/(\theta_*+\delta)^{s+1}\leq L.
\]
Both densities are in the same fixed class. For $y<0$ their observed densities are equal. Their atom integrals are equal. For every $y>0$, $y+t_2>t_2$ and
\[
 f_2(y+t_2)=f_1(y+t_2-d\delta)=f_1(y+t_1).
\]
Thus all parts of the full law agree. Mixing both laws with the same $\gamma\delta_0$ preserves equality.

Under the common product law, $|\widehat\theta-\theta_1|+|\widehat\theta-\theta_2|\geq\delta$ pointwise. At least one of the expected errors is at least $\delta/2$. For a confidence set covering both models at level $1-\alpha$, the probability of including both values is at least $1-2\alpha$. Its expected diameter is at least $(1-2\alpha)\delta$. These conclusions hold for every sample size.
\end{proof}

\subsection{Distances between complete observation laws}
We use $H^2(P,Q)=\int(\sqrt{\mathrm dP}-\sqrt{\mathrm dQ})^2$ and affinity $\mathfrak a(P,Q)=\int\sqrt{\mathrm dP\,\mathrm dQ}=1-H^2(P,Q)/2$. The dominating measure is Lebesgue measure plus a unit atom at zero.

\begin{lemma}[Full-law stochastic distance]\label{lem:hellinger}
For the known density in \eqref{eq:plateau}, $0<d\leq d_0$, and known $\gamma\leq\bar\gamma$, the full observed laws satisfy \eqref{eq:Hbound}, uniformly over $\theta_1,\theta_2\in\ThetaS$.
\end{lemma}
\begin{proof}
Shrink $d_0$ so that $d_0T<1$. In the plateau submodel the atom probability is $r_j=\gamma+(1-\gamma)vd\theta_j$. Since $\theta_j\geq a$ and $1-\gamma\geq1-\bar\gamma$, $r_j\geq K^{-1}(\gamma+d)$. Therefore
\[
 (\sqrt{r_1}-\sqrt{r_2})^2
 =\frac{(r_1-r_2)^2}{(\sqrt{r_1}+\sqrt{r_2})^2}
 \leq K\frac{d^2(\theta_1-\theta_2)^2}{\gamma+d}.
\]
The negative densities coincide. Write $w=\sqrt{f_1}$. This function is absolutely continuous with weak derivative in $L^2(\R)$. Direct integration gives $\int(w')^2=\pi/16$. For every real $r$, the fundamental theorem of calculus and Cauchy--Schwarz give
\[
 |w(x+r)-w(x)|^2\leq |r|\int_{\min(x,x+r)}^{\max(x,x+r)}|w'(u)|^2\dd u.
\]
Integrating in $x$ and using Tonelli gives the translation bound $\int|w(x+r)-w(x)|^2\dd x\leq r^2\int(w')^2$. The positive-side contribution to squared Hellinger distance is consequently at most $(1-\gamma)d^2(\theta_1-\theta_2)^2\pi/16$. Since $\gamma+d$ is uniformly bounded, this is bounded by the same order as the atom term. Adding contributions proves the lemma for complete laws.
\end{proof}

Write $\TV(P,Q)=\tfrac12\int|p-q|$ for total variation distance, using a common dominating measure. For completeness, the product and testing inequalities used below can be verified directly. Product affinity equals $\mathfrak a(P,Q)^n$. By Cauchy--Schwarz applied to $(\sqrt p-\sqrt q)(\sqrt p+\sqrt q)$,
\[
 \TV(P,Q)\leq\sqrt{1-\mathfrak a(P,Q)^2},\qquad
 \TV(P^{\otimes n},Q^{\otimes n})\leq\sqrt{nH^2(P,Q)}.
\]
The last inequality follows from $1-(1-H^2/2)^{2n}\leq nH^2$. If two parameters differ by $\Delta$, integrate the pointwise loss triangle inequality against the common measure $\min(p,q)$. It gives
\begin{equation}\label{eq:testingloss}
 \max_j\E_j|\widehat\theta-\theta_j|
 \geq \frac\Delta2(1-\TV(P_1,P_2)).
\end{equation}
For honest sets, under the first law the second parameter is included with probability at least $1-\alpha-\TV(P_1,P_2)$. A further union bound gives
\begin{equation}\label{eq:testingdiam}
 \E_1\diam(C)\geq\Delta[1-2\alpha-\TV(P_1,P_2)]
\end{equation}
when the bracket is positive.

\subsection{Risk bounds and the design balance}
\begin{proof}[Proof of \cref{thm:minimax}]
For the lower bound choose an interior $\theta_*$ and a second parameter separated by
\[
 \Delta=\min\{\Delta_0,\xi\sqrt{(\gamma+d)/(nd^2)}\},
\]
where $\Delta_0$ keeps the pair inside $\ThetaS$. By \cref{lem:hellinger}, $nH^2\leq K\xi^2$. Choose $\xi$ small so the product total variation is at most $1/2$. Equation \eqref{eq:testingloss} then gives a constant times $\min\{1,\sqrt{(\gamma+d)/(nd^2)}\}$. The equivalence construction supplies a separate lower bound $\kappa d^s/2$. For $d_0\leq1$, the maximum of these two terms is bounded below by a positive multiple of $\rho_{n,d,\gamma}$.

For the upper bound let $h=d$, $\mu=q/d$, and $\overline\mu=\max(\widehat\mu_d,c_*/2)$. Since $\mu\geq c_*$, $|\overline\mu-\mu|\leq|\widehat\mu_d-\mu|$. Also $(p-\gamma)/d\leq CT$. Thus
\[
 \left|\frac{\widehat p-\gamma}{d\overline\mu}-\theta_h\right|
 \leq\frac2{c_*d}|\widehat p-p|
 +\frac{2CT}{c_*^2}|\widehat\mu_d-\mu|.
\]
Binomial marginal variances and Jensen imply
\[
 \E|\widehat p-p|\leq\sqrt{p(1-p)/n}\leq K\sqrt{(\gamma+d)/n},\quad
 \E|\widehat\mu_d-\mu|\leq\sqrt{q(1-q)/(nd^2)}\leq K/\sqrt{nd}.
\]
The second stochastic term is no larger in order than the first after structural scaling. Equation \eqref{eq:pseudobias} contributes $Kd^s$. Projection onto $\ThetaS$ cannot increase error relative to the true parameter and bounds error by $T-a$. This proves the clipped upper bound. At $d=0$, the two endpoint parameters have the same law and their loss sum is at least $T-a$; the midpoint achieves worst-case loss $(T-a)/2$.
\end{proof}

\begin{proof}[Proof of \cref{cor:balance}]
The upper risk vanishes under the two stated conditions. A bounded $\mathcal A$ leaves the stochastic lower bound positive, and a fixed positive $d\leq d_0$ leaves the equivalence lower bound positive. For the balance, $nd^{2s+2}/(\gamma+d)$ is strictly increasing from zero to infinity: its log derivative is $(2s+2)/d-1/(\gamma+d)>0$. Thus the equation has a unique positive solution. Put $r_n=n^{-1/(2s+1)}$ and $d=r_n x$. Then the equation becomes $x^{2s+2}=\gamma/r_n+x$. If $\gamma/r_n$ is bounded, $x$ is bounded above and away from zero. If it diverges, $x\asymp(\gamma/r_n)^{1/(2s+2)}$, with $x/(\gamma/r_n)\to0$. Substituting gives the stated kink and risk orders. The solution eventually lies in the local design range. The sum of a decreasing stochastic term and an increasing ambiguity term is within fixed factors of its balanced order; hence this is also the order of the best achievable precision over local choices of $d$.
\end{proof}

\subsection{Conventional asymptotics}
\begin{proof}[Proof of \cref{thm:limits}(i)]
Write $A_i=\1\{Y_i=0\}$, $B_i=\1\{-h\leq Y_i<0\}$, $\beta=(p-\gamma)/q$, and $Z_i=A_i-\gamma-\beta B_i$. The cells are disjoint, $\E Z_i=0$, and
\[
 \operatorname{Var}(Z_i)=p(1-p)+\beta^2q(1-q)+2\beta pq
 =p-\gamma^2+(p-\gamma)^2/q=V_h.
\]
Since $p-\gamma^2=\gamma(1-\gamma)+(p-\gamma)\geq(1-\bar\gamma)p$, and $\beta\geq0$, $V_h$ bounds below fixed multiples of $p$ and $\beta^2q$. Moreover $\gamma\leq p$. Bounding centered third absolute moments using the three cell values gives
\[
 \frac{\E|Z_i|^3}{\sqrt n V_h^{3/2}}
 \leq K\{(np)^{-1/2}+(nq)^{-1/2}\}.
\]
One may instead verify Lindeberg by bounding the maximal summand size by $1+\gamma+\beta$ and dividing by $\sqrt{nV_h}$. The triangular iid central limit theorem yields $\sum_i Z_i/\sqrt{nV_h}\Rightarrow N(0,1)$; the displayed bound also gives uniform convergence when both expected counts diverge uniformly.

The exact ratio identity on $\widehat q>0$ is
\[
 \widehat\theta_h-\theta_h
 =\frac{\widehat p-\gamma-\beta\widehat q}{d\widehat q/h}
 =\frac{h}{d\widehat q}\frac1n\sum_i Z_i.
\]
Chebyshev gives $\widehat q/q\to_{\Prob}1$ uniformly if $nq\to\infty$, so the zero-denominator event has probability tending to zero. This proves the normal conclusion with variance $v_h$.

For studentization, let $e_p=\widehat p-p$ and $e_q=\widehat q-q$. Relative binomial errors satisfy $e_p/p=O_\Prob((np)^{-1/2})$ and $e_q/q=O_\Prob((nq)^{-1/2})$. The change in the first numerator term is $e_p$. The linear change in $(p-\gamma)^2/q$ has magnitude at most a constant times $|p-\gamma||e_p|/q+(p-\gamma)^2|e_q|/q^2$, apart from $e_p^2/q$ and products with $e_q/q$. Since $V_h\geq K^{-1}p$ and $V_h\geq(p-\gamma)^2/q$, these changes divided by $V_h$ are bounded in probability by
\[
 K\{(np)^{-1/2}+(nq)^{-1/2}+(nq)^{-1}\}=o_\Prob(1).
\]
The denominator ratio is $(\widehat q/q)^2\to1$. Thus $\widehat v_h/v_h\to1$. Adding a population centering error negligible relative to $\sqrt{v_h}$ proves structural centering. The bridge supplies the stated sufficient condition. Bounds used throughout are uniform over the fixed class.
\end{proof}

\begin{proof}[Proof of \cref{thm:limits}(ii)--(iii)]
For $\gamma=0$, local smoothness gives $p=d\theta f(0)+O(d^{s+1})$. Hence $np\to\lambda=\tau\theta f_*$. The probability generating function of $N_0$ is $(1+p(z-1))^n$, which converges to $\exp\{\lambda(z-1)\}$ for each fixed $z\in[0,1]$. This proves its Poisson limit. The bin estimator has stochastic variance $O((nh)^{-1})$ and bias $O(h^s)$ for $f(0)$, so $\widehat\mu_h\to_{\Prob}f_*$. Slutsky applied to $N_0/(nd\widehat\mu_h)$ gives the ratio limit. When $nd\to0$, $\Prob(N_0>0)\leq np\to0$, and the nonzero-denominator event tends to one. Thus the ratio tends to zero in probability, regardless of its possible rare large values.

When $n\gamma\to\zeta$, $\gamma\to0$ and the total atom mean converges to $\zeta+\lambda$. Its generating function has the same limit with that mean. The density estimator still tends to $f_*$. The exact ratio is $(N_0-n\gamma)/(nd\widehat\mu_h)$, giving (iii). This argument proves limits of these statistics; it does not by itself identify a full-sample experiment.
\end{proof}

\begin{proof}[Proof of \cref{thm:wald}]
At $\gamma=0$, the event $N_0=0$, $N_->0$ yields $\widehat\theta_h=0$ and $\widehat v_h=0$. It excludes every $\theta_n\geq a$. The event $N_-=0$ has probability at most $\exp(-nq)\to0$. Meanwhile $\Prob(N_0=0)=(1-p)^n\to e^{-\lambda}$. Thus $\Prob(\theta_n\in W_n)\leq1-\Prob(N_0=0,N_->0)$ has the stated limiting upper bound. Choosing a fixed $\lambda<-\log\alpha$ provides an admissible sequence violating uniform nominal coverage.
\end{proof}

\subsection{Exact coverage and unconditional diameter}
On the simultaneous probability-coverage event, $p_U\geq p\geq\gamma$ also verifies the nonnegative behavioral-mass constraint in \eqref{eq:CS}. Thus adding this necessary feasibility condition preserves the coverage argument below.
\begin{proof}[Proof of \cref{thm:coverage}]
Each count has a binomial marginal. Exact inversion guarantees $\Prob(p\in[p_L,p_U])\geq1-\alpha_p$ and $\Prob(q\in[q_L,q_U])\geq1-\alpha_q$. Their simultaneous event has probability at least $1-\alpha$, despite dependence. Every $\vartheta\in I(P)$ has some compatible density in the same class, so the bridge applies using the same population $p,q,d,\gamma$. On the simultaneous event its population inequalities imply both sample inequalities in \eqref{eq:CS}. This proves inclusion of the whole identified set on one event. At $d=0$ the inequalities reduce to $p_L\leq\gamma\leq p_U$, proving the stated special cases. All bounds hold for every law and finite $n$, so taking an infimum introduces no limit issue.
\end{proof}

\begin{lemma}[Widths of exact binomial intervals]\label{lem:cpwidth}
For $K\sim\operatorname{Bin}(n,u)$, $x=K/n$, fixed error $\beta$, and $\ell=\log(2/\beta)$, exact endpoints obey
\[
 \ell_\beta(K)\geq\max\{0,x-\sqrt{2x\ell/n}\},\qquad
 u_\beta(K)\leq\min\{1,x+\sqrt{2x\ell/n}+2\ell/n\}.
\]
Consequently their expected width is at most $2\sqrt{2u\ell/n}+2\ell/n$.
\end{lemma}
\begin{proof}
The binomial Chernoff tail bound at a proposed probability $v$ is $\exp[-nD(x\Vert v)]$, with $D$ the binary relative entropy. Integrating $\partial_vD(x\Vert v)=(v-x)/(v(1-v))$ gives $D(x\Vert v)\geq(x-v)^2/(2x)$ for $v<x$, and $D(x\Vert v)\geq(v-x)^2/(2v)$ for $v>x$. The proposed lower endpoint makes the first exponent at least $\ell$. For the upper endpoint put $a=\ell/n$ and $v=x+\sqrt{2xa}+2a$. Then $(v-x)^2\geq2av$, so the upper-tail exponent is at least $\ell$. These values conservatively bracket the exact tail inversions; clipping at zero and one handles boundaries. Jensen's inequality $\E\sqrt{x}\leq\sqrt u$ gives the expected-width bound. All bounds concern marginal intervals and require no independence from another count.
\end{proof}

\begin{proof}[Proof of \cref{thm:diameter}]
For the lower bound use the same nearby full-law pair as in \cref{thm:minimax}, but choose $\xi$ so product total variation is below $(1-2\alpha)/2$. Equation \eqref{eq:testingdiam} gives a constant times the clipped stochastic rate. The observational-equivalence pair gives $(1-2\alpha)\kappa d^s$. Their maximum bounds below $K_3\rho$. At $d=0$ the two parameter-space endpoints have the same law, giving a positive constant lower bound.

For the upper bound, combine the two acceptance inequalities at arbitrary $\vartheta_1<\vartheta_2$. The terms involving $\gamma$ cancel, giving \eqref{eq:diamdet}; use $b_\gamma(t,h)\leq Lt(t+h)^s$. On the event $G=\{q_L/d\geq c_*/2\}$, with $h=d$, the diameter is at most
\[
 \frac2{c_*}\left\{\frac{p_U-p_L}{d}+T\frac{q_U-q_L}{d}+2LT(dT+d)^s\right\}.
\]
This statement holds for every accepted pair, whether or not the true parameter is accepted, and whether or not the set is connected. Its empty-set diameter convention also satisfies the bound.

Let $\ell_q=\log(2/\alpha_q)$. If $nd\geq24\ell_q/c_*$, $\widehat q\geq3c_*d/4$ implies $q_L/d\geq c_*/2$ by \cref{lem:cpwidth}. The lower bounding function $x-\sqrt{2x\ell_q/n}$ is increasing above this threshold, and its square-root term there is at most $c_*d/4$. Because $q\geq c_*d$, a multiplicative binomial Chernoff bound gives $\Prob(G^c)\leq e^{-c_*nd/32}$.

Apply \cref{lem:cpwidth} using $p\leq K(\gamma+d)$ and $q\leq Cd$. The unconditional expectation of the displayed good-event bound is at most
\[
 K\left\{\sqrt{\frac{\gamma+d}{nd^2}}+(nd)^{-1/2}+(nd)^{-1}+d^s\right\}.
\]
On $G^c$ diameter is at most $T-a$. For $nd$ above the stated fixed threshold, the exponential exceptional probability and $(nd)^{-1}$ are bounded by constants times $(nd)^{-1/2}$, which is dominated by the principal stochastic term. For smaller $nd$, $\rho$ is bounded below by a positive constant and the bounded parameter-space diameter suffices. Finally clip the whole expectation by $T-a$. This proves the unconditional upper rate. At $d=0$ the bounded diameter supplies the upper constant order. Every constant is independent of $\gamma,n,d$ in their stated ranges.
\end{proof}

\begin{proof}[Proof of \cref{prop:control}]
Form exact marginal intervals for the two treatment counts and the control zero count with errors totaling $\alpha$. On their simultaneous coverage event, the true pair $(\theta,\gamma)$ belongs to the joint acceptance set by the bridge. Projection includes its true elasticity. The union bound gives probability at least $1-\alpha$. Independence of the control sample justifies its stated sampling model; the coverage step itself uses only the valid three marginals. No claim about the rate of this projected set is used.
\end{proof}

\section{A likelihood-ratio limit and additional boundary checks}
\begin{theorem}[Finite-parameter Poisson likelihood limit]\label{thm:poissonlr}
Let the counterfactual density be known and equal to \eqref{eq:plateau}, $\gamma=0$, and $d_n>0$ with $nd_n\to\tau\in[0,\infty)$. Fix $\theta_0\in\ThetaS$ and any finite collection $\theta_1,\ldots,\theta_k\in\ThetaS$. Under the full true product law, their joint log likelihood ratios converge to
\[
 \left(N\log\frac{\theta_j}{\theta_0}-\tau v(\theta_j-\theta_0)\right)_{j=1}^k,
 \qquad N\sim\Pois(\tau v\theta_0),\quad v=1/4.
\]
For $\tau=0$ this vector is zero almost surely. The statement is convergence for every fixed finite collection, not uniform deficiency-distance equivalence over the semiparametric class.
\end{theorem}
\begin{proof}
For large $n$, $d_nT<1$, so the atom probability is exactly $vd_n\theta$. Let $\ell(x)=\log f_1(x)$. On $x\geq0$, $\ell'(x)=-(x-1)_+/\sigma^2$ is Lipschitz with constant $1/\sigma^2$. The negative-side likelihood ratio is one. The atom contribution is $N_0\log(\theta_j/\theta_0)$. Taylor's formula with a Lipschitz derivative gives the positive-side contribution
\[
 d_n(\theta_j-\theta_0)\sum_i S_i+R_{n,j},\quad
 S_i=\1\{Y_i>0\}\ell'(Y_i+d_n\theta_0),\quad
 |R_{n,j}|\leq nd_n^2(\theta_j-\theta_0)^2/(2\sigma^2).
\]
Under the true law,
\[
 \E S_i=\int_{d_n\theta_0}^\infty f_1(x)\ell'(x)\dd x
 =\int_{d_n\theta_0}^\infty f_1'(x)\dd x=-v.
\]
The second moment is bounded by a finite Gaussian-tail integral, uniformly in $n$. Thus $d_n\sum_i(S_i+v)=O_\Prob(d_n\sqrt n)=o_\Prob(1)$, and $nd_n^2\to0$. The atom count converges to $\Pois(\tau v\theta_0)$ as proved above. The same count occurs for all alternatives; the finite maximum of their remainders tends to zero. Joint convergence follows by Slutsky. At $\tau=0$ every term vanishes in probability. No claim about an infinite collection of alternatives or nuisance densities follows from this finite-dimensional proof.
\end{proof}

\subsection{A check with a Gaussian counterfactual}
For a known centered normal density with standard deviation $\sigma$, write $t=d\theta_1$ and $u=d\theta_2$. The complete-law affinity when $\gamma=0$ is
\[
 \frac12+
 \sqrt{[\Phi(t/\sigma)-1/2][\Phi(u/\sigma)-1/2]}+
 e^{-(t-u)^2/(8\sigma^2)}\overline\Phi((t+u)/(2\sigma)).
\]
The negative half-line gives $1/2$; the atom gives the square-root product; completing the square gives the upper integral. For fixed distinct parameters its squared Hellinger distance equals
\[
 \phi_\sigma(0)d(\sqrt{\theta_1}-\sqrt{\theta_2})^2+O(d^2).
\]
This provides a symbolic and numerical check of the rare-atom order. The smooth plateau submodel in the main proof is preferred because it satisfies every positive local H\"older radius without a further normal-density inclusion condition.

\subsection{The meaning of nonidentification at a fixed kink}
The population identified set in \eqref{eq:identifiedset} depends on the permitted densities. Setting $L=0$ on the whole local interval changes the model: the density is constant there, the ambiguity construction is unavailable, and the local-flat ratio identifies elasticity for $d>0$. Our minimax lower bound explicitly assumes $L>0$. Likewise, assuming a globally known normal density gives oracle identification; the simulations use normal data but do not give that parametric information to the procedures. These changes cannot be treated as mere tuning choices within the same theorem.

The positive parameter bound $a>0$ also matters. If elasticities approach zero, the atom may vanish faster than the policy size and the boundary can affect both normal and Poisson limits. If $f(0)$ approaches zero, the pool of potential movers changes. If $\gamma\to1$, the accurately recorded share vanishes. None of these sequences belongs to the fixed class, so they are not implicit extensions of the uniformity claims.

\subsection{Background atoms and weak normal limits}
For fixed $\gamma\in(0,1)$ and $d_n=\kappa/\sqrt n$, the atom estimator obeys a binomial central limit theorem with variance approaching $\gamma(1-\gamma)/n$. The population net atom equals $(1-\gamma)d_n\theta f_*+o(d_n)$. Under $h_n\to0$, $nh_n\to\infty$, and bounded $d_n/h_n$, the density estimate tends to $(1-\gamma)f_*$, its contribution to the ratio vanishes, and its deterministic bridge discrepancy tends to zero. This proves \eqref{eq:bg-normal} for the count-based ratio. The full-law continuous shift is of order $d_n$, so its cumulative known-density information can remain of order $nd_n^2$. We therefore do not identify the complete experiment with the single normal count statistic in this regime.

\section{Computational implementation and reproducibility}
\subsection{Probability calculations and component roots}
For $X\sim N(\mu,1)$, let $\Phi$ denote the standard normal distribution function. The exact cell probabilities are
\[
 B=\Phi(d\theta-\mu)-\Phi(-\mu),\qquad
 Q=\Phi(-\mu)-\Phi(-h-\mu).
\]
With a background mixture they become $p=\gamma+(1-\gamma)B$ and $q=(1-\gamma)Q$. The three probabilities $(p,q,1-p-q)$ define the exact multinomial shortcut used in the experiments. Seeds are fixed by experiment and cell index. Code also checks the shortcut against directly simulated optimizing choices.

For $s=1$ and a valid scaled-density Lipschitz bound $L_g$, put $t=d\vartheta$ and $b_U=q_U/h+L_gh/2$. The upper-envelope constraint requires
\[
 t\geq\frac{2\max(0,p_L-\gamma)}{b_U+\sqrt{b_U^2+2L_g\max(0,p_L-\gamma)}}.
\]
If $p_U-\gamma<0$, the set is empty. Otherwise let $b_L=q_L/h-L_gh/2$. If $b_L>0$ and $b_L^2>2L_g(p_U-\gamma)$, the lower-envelope constraint accepts $t$ below the smaller root or above the larger root:
\[
 r_- =\frac{2(p_U-\gamma)}{b_L+\sqrt{b_L^2-2L_g(p_U-\gamma)}},\qquad
 r_+=\frac{b_L+\sqrt{b_L^2-2L_g(p_U-\gamma)}}{L_g}.
\]
In the other cases it accepts every nonnegative $t$. Intersect all constraints with $[da,dT]$ and divide by $d$. The implementation handles $L_g=0$ separately with linear inequalities and handles $d=0$ before division. Equality endpoints are retained. Total component length differs from diameter when the set is disconnected.

\subsection{Validation and numerical provenance}
Tests check recovery in a locally flat strong-count benchmark, exact marginal coverage by enumerating count outcomes, null coverage in simulated structural data, weak and zero-kink edge cases, fixed-seed reproduction, and the analytic component solver against direct inequalities. Independent symbolic checks verify the multinomial covariance cancellation, including the subtraction of $\gamma^2$ in the background model. Independent quadrature checks the bridge, complete-law affinity, smooth-tail normalization, and observational-equivalence mass. Such checks support implementation and detect errors; they are not substitutes for the proofs.

Generated simulation tables contain no manually inserted results. A provenance file records the summary-file hash, selected cell identifiers, macro values used in the text, and rounding. Replication-level arrays contain both counts, point estimates, standard errors, coverage indicators, set lengths, diameters, and component counts. Summary files include all design inputs and seeds. The manuscript's numerical statements are produced from these summaries. The analytical rate plot is explicitly labeled as a rate expression with constants suppressed.

\subsection{Replication sequence}
The replication program runs mathematical and estimator tests, executes all three experiments, generates tables and figures, checks numerical provenance and verified citation keys, and compiles the multi-file manuscript. It requires no graphics processing unit and uses vectorized multinomial draws with bounded memory. TeX compilation uses relative source paths. The project records package versions and build diagnostics. A source-only archive is rebuilt in an isolated directory, and its regenerated document is checked separately from the working-directory build.

The source package contains only the main file, sections, appendices, bibliography, generated table inputs, figures, and the compiled bibliography needed by the TeX workflow. Research logs, referee reports, locally downloaded literature, raw simulations, and credentials are excluded. An arXiv-compatible source archive does not imply a completed submission or an announced identifier.

\bibliographystyle{plainnat}
\bibliography{references}

\begin{thebibliography}{21}
\providecommand{\natexlab}[1]{#1}
\providecommand{\url}[1]{\texttt{#1}}
\expandafter\ifx\csname urlstyle\endcsname\relax
  \providecommand{\doi}[1]{doi: #1}\else
  \providecommand{\doi}{doi: \begingroup \urlstyle{rm}\Url}\fi

\bibitem[Andrews and Cheng(2012)]{andrews2012}
Donald W.~K. Andrews and Xu~Cheng.
\newblock {Estimation and Inference With Weak, Semi-Strong, and Strong
  Identification}.
\newblock \emph{Econometrica}, 80\penalty0 (5):\penalty0 2153--2211, 2012.
\newblock \doi{10.3982/ECTA9456}.

\bibitem[Armstrong and Koles{\'a}r(2018)]{armstrong2018}
Timothy~B. Armstrong and Michal Koles{\'a}r.
\newblock {Optimal Inference in a Class of Regression Models}.
\newblock \emph{Econometrica}, 86\penalty0 (2):\penalty0 655--683, 2018.
\newblock \doi{10.3982/ECTA14434}.

\bibitem[Bertanha et~al.(2023)Bertanha, McCallum, and Seegert]{bertanha2023}
Marinho Bertanha, Andrew~H. McCallum, and Nathan Seegert.
\newblock {Better bunching, nicer notching}.
\newblock \emph{Journal of Econometrics}, 237\penalty0 (2):\penalty0 105512,
  2023.
\newblock \doi{10.1016/j.jeconom.2023.105512}.

\bibitem[Blomquist et~al.(2021)Blomquist, Newey, Kumar, and
  Liang]{blomquist2021}
S{\"o}ren Blomquist, Whitney~K. Newey, Anil Kumar, and Che-Yuan Liang.
\newblock {On Bunching and Identification of the Taxable Income Elasticity}.
\newblock \emph{Journal of Political Economy}, 129\penalty0 (8):\penalty0
  2320--2343, 2021.
\newblock \doi{10.1086/714446}.

\bibitem[Caetano et~al.(2025)Caetano, Caetano, Goff, and Nielsen]{caetano2025}
Carolina Caetano, Gregorio Caetano, Leonard Goff, and Eric Nielsen.
\newblock {Identification of Causal Effects with a Bunching Design}.
\newblock arXiv:2507.05210v1, 2025.
\newblock URL \url{https://arxiv.org/abs/2507.05210v1}.
\newblock Version 1, July 7, 2025.

\bibitem[Cattaneo et~al.(2020)Cattaneo, Jansson, and Ma]{cattaneo2020}
Matias~D. Cattaneo, Michael Jansson, and Xinwei Ma.
\newblock {Simple Local Polynomial Density Estimators}.
\newblock \emph{Journal of the American Statistical Association}, 115\penalty0
  (531):\penalty0 1449--1455, 2020.
\newblock \doi{10.1080/01621459.2019.1635480}.

\bibitem[Chetty et~al.(2011)Chetty, Friedman, Olsen, and
  Pistaferri]{chetty2011}
R.~Chetty, J.~N. Friedman, T.~Olsen, and L.~Pistaferri.
\newblock {Adjustment Costs, Firm Responses, and Micro vs. Macro Labor Supply
  Elasticities: Evidence from Danish Tax Records}.
\newblock \emph{The Quarterly Journal of Economics}, 126\penalty0 (2):\penalty0
  749--804, 2011.
\newblock \doi{10.1093/qje/qjr013}.

\bibitem[Clopper and Pearson(1934)]{clopper1934}
C.~J. Clopper and E.~S. Pearson.
\newblock {The Use of Confidence or Fiducial Limits Illustrated in the Case of
  the Binomial}.
\newblock \emph{Biometrika}, 26\penalty0 (4):\penalty0 404--413, 1934.
\newblock \doi{10.1093/biomet/26.4.404}.

\bibitem[Dufour(1997)]{dufour1997}
Jean-Marie Dufour.
\newblock {Some Impossibility Theorems in Econometrics With Applications to
  Structural and Dynamic Models}.
\newblock \emph{Econometrica}, 65\penalty0 (6):\penalty0 1365--1387, 1997.
\newblock \doi{10.2307/2171740}.

\bibitem[Dutz et~al.(2026)Dutz, H{\aa}varstein, Mogstad, and
  Torgovitsky]{dutz2026}
Deniz Dutz, Morten H{\aa}varstein, Magne Mogstad, and Alexander Torgovitsky.
\newblock {Identification and Estimation of Labor Supply Elasticities from
  Kinked Budget Sets}.
\newblock BFI Working Paper 2026-53, 2026.
\newblock URL
  \url{https://bfi.uchicago.edu/wp-content/uploads/2026/04/BFI_WP_2026-53.pdf}.
\newblock PDF cover April 1, 2026; institutional posting April 2, 2026.

\bibitem[Goff(2021)]{goff2021}
Leonard Goff.
\newblock {Treatment Effects in Bunching Designs: The Impact of the Federal
  Overtime Rule on Hours}.
\newblock Working paper, 2021.
\newblock URL
  \url{https://econ.columbia.edu/wp-content/uploads/sites/32/2020/09/goff_jmp_overtime_kink.pdf}.
\newblock PDF cover: January 17, 2021; superseded version, cited for background
  bunchers.

\bibitem[Goff(2026)]{goff2026}
Leonard Goff.
\newblock {Partial Identification from Bunching at Kinks and Notches}.
\newblock arXiv:2205.10310v5, 2026.
\newblock URL \url{https://arxiv.org/abs/2205.10310v5}.
\newblock Version 5, revised September 9, 2026.

\bibitem[Kaji(2021)]{kaji2021}
Tetsuya Kaji.
\newblock {Theory of Weak Identification in Semiparametric Models}.
\newblock \emph{Econometrica}, 89\penalty0 (2):\penalty0 733--763, 2021.
\newblock \doi{10.3982/ECTA16413}.

\bibitem[Kleven and Waseem(2013)]{kleven2013}
Henrik~J. Kleven and Mazhar Waseem.
\newblock {Using Notches to Uncover Optimization Frictions and Structural
  Elasticities: Theory and Evidence from Pakistan}.
\newblock \emph{The Quarterly Journal of Economics}, 128\penalty0 (2):\penalty0
  669--723, 2013.
\newblock \doi{10.1093/qje/qjt004}.

\bibitem[Kleven(2016)]{kleven2016}
Henrik~Jacobsen Kleven.
\newblock {Bunching}.
\newblock \emph{Annual Review of Economics}, 8\penalty0 (1):\penalty0 435--464,
  2016.
\newblock \doi{10.1146/annurev-economics-080315-015234}.

\bibitem[Moreira(2003)]{moreira2003}
Marcelo~J. Moreira.
\newblock {A Conditional Likelihood Ratio Test for Structural Models}.
\newblock \emph{Econometrica}, 71\penalty0 (4):\penalty0 1027--1048, 2003.
\newblock \doi{10.1111/1468-0262.00438}.

\bibitem[Saez(2010)]{saez2010}
Emmanuel Saez.
\newblock {Do Taxpayers Bunch at Kink Points?}
\newblock \emph{American Economic Journal: Economic Policy}, 2\penalty0
  (3):\penalty0 180--212, 2010.
\newblock \doi{10.1257/pol.2.3.180}.

\bibitem[Song(2025)]{song2025}
Myunghyun Song.
\newblock {Identification and Inference in General Bunching Designs}.
\newblock arXiv:2411.03625v3, 2025.
\newblock URL \url{https://arxiv.org/abs/2411.03625v3}.
\newblock Version 3, revised February 6, 2025; locked PDF cover February 7,
  2025.

\bibitem[Staiger and Stock(1997)]{staiger1997}
Douglas Staiger and James~H. Stock.
\newblock {Instrumental Variables Regression with Weak Instruments}.
\newblock \emph{Econometrica}, 65\penalty0 (3):\penalty0 557--586, 1997.
\newblock \doi{10.2307/2171753}.

\bibitem[Stock and Wright(2000)]{stock2000}
James~H. Stock and Jonathan~H. Wright.
\newblock {GMM with Weak Identification}.
\newblock \emph{Econometrica}, 68\penalty0 (5):\penalty0 1055--1096, 2000.
\newblock \doi{10.1111/1468-0262.00151}.

\bibitem[Wang et~al.(2025)Wang, Chan, and Ye]{wang2025}
Rui Wang, Kwun Chuen~Gary Chan, and Ting Ye.
\newblock {GMM with Many Weak Moment Conditions and Nuisance Parameters:
  General Theory and Applications to Causal Inference}.
\newblock arXiv:2505.07295v3, 2025.
\newblock URL \url{https://arxiv.org/abs/2505.07295v3}.
\newblock Version 3, revised November 10, 2025.

\end{thebibliography}
\end{document}